\def\makeSM{1}
\documentclass[reprint,twocolumn,superscriptaddress,showpacs,floatfix,longbibliography]{revtex4-1}
\usepackage{color}
\usepackage{mathtools}
\usepackage{bm}        \usepackage{amssymb}   \usepackage{amsmath}
\usepackage{wasysym}
\usepackage[colorlinks=true,citecolor=blue,linkcolor=magenta, breaklinks=true]{hyperref}
\usepackage{graphicx}
\usepackage{braket}
\usepackage{natbib}
\usepackage [latin1]{inputenc}
\usepackage{siunitx}
\usepackage{blkarray, bigstrut}
\usepackage[dvipsnames]{xcolor}
\usepackage{comment}
\usepackage{physics}
\graphicspath{{./Figures/}}

\usepackage{tikz}
\newcommand{\foo}[1]{\begin{tikzpicture}[#1]\draw (0,0) -- (1ex,1ex);\draw (0.5ex,0) -- (1.3ex,0.8ex);\end{tikzpicture}}

\ifdefined\makeSM
\usepackage{xpatch}

\makeatletter
\patchcmd{\@ssect@ltx}
    {\addcontentsline{toc}{#1}{\protect\numberline{}#8}}
    {}
    {}
    {}
\makeatother
\fi

\newcommand{\bea}{\begin{eqnarray}}
\newcommand{\eea}{\end{eqnarray}}
\newcommand{\nn}{\langle ij \rangle}
\newcommand{\nnn}{\langle\langle ij \rangle\rangle}
\newcommand{\bsigma}{\boldsymbol{\sigma}}
\newcommand{\bS}{\mathbf{S}}
\newcommand{\bb}{\mathbf{b}}
\newcommand{\br}{\mathbf{r}}
\newcommand{\hatz}{\hat{z}}
\newcommand{\ba}{\mathbf{a}}
\newcommand{\bs}{\mathbf{s}}
\newcommand{\bp}{\mathbf{p}}
\newcommand{\bn}{\mathbf{n}}
\newcommand{\bG}{\mathbf{G}}
\newcommand{\bOm}{\mathbf{\Omega}}

\newcommand{\mb}{\mathfrak{b}}
\newcommand\bbone{\ensuremath{\mathbbm{1}}}
\newcommand{\ul}{\underline}
\newcommand{\vl}{v_{_L}}
\newcommand{\vc}{\mathbf}
\newcommand{\be}{\begin{equation}}
\newcommand{\ee}{\end{equation}}
\newcommand{\bk}{{{\bf{k}}}}
\newcommand{\bK}{{{\bf{K}}}}
\newcommand{\cE}{{{\cal E}}}
\newcommand{\bQ}{{{\bf{Q}}}}
\newcommand{\bg}{{{\bf{g}}}}
\newcommand{\hbr}{{\hat{\bf{r}}}}
\newcommand{\bR}{{{\bf{R}}}}
\newcommand{\bq}{{\bf{q}}}
\newcommand{\hx}{{\hat{x}}}
\newcommand{\hy}{{\hat{y}}}
\newcommand{\ha}{{\hat{a}}}
\newcommand{\hb}{{\hat{b}}}
\newcommand{\hc}{{\hat{c}}}
\newcommand{\hd}{{\hat{\delta}}}
\newcommand{\beal}{\begin{align}}
\newcommand{\eeal}{\end{align}}
\newcommand{\ra}{\rangle}
\newcommand{\la}{\langle}
\renewcommand{\tt}{{\tilde{t}}}
\newcommand{\upa}{\uparrow}
\newcommand{\dna}{\downarrow}
\newcommand{\vS}{\vec{S}}
\newcommand{\dg}{{\dagger}}
\newcommand{\pdg}{{\phantom\dagger}}
\newcommand{\tphi}{{\tilde\phi}}
\newcommand{\cf}{{\cal F}}
\newcommand{\ca}{{\cal A}}
\renewcommand{\ni}{\noindent}
\def\l{\ell}
\newcommand{\trs}{\mathcal{T}}
\newcommand{\bdel}{\boldsymbol{\delta}}
\renewcommand{\bm}{\mathbf{m}}

\def\Jt{\mathbf{J}}
\def\S{\tilde{\mathbf{S}}}
\def\Sc{\tilde{S}}
\def\h{\mathbf{h}}

\renewcommand{\vec}[1]{\mathbf{#1}}
\def\vS{\vec{S}}

\newcommand{\btjstrw}{\mathrel{{\rotatebox[origin=c]{90}
{$\bowtie$}}\kern-0.18em\raisebox{-.95ex}{$\bullet$}
\kern-0.5em\raisebox{.97ex}{$\bullet$}
\kern-1.12em\raisebox{.97ex}{$\bullet$}
\kern-0.52em\raisebox{-.95ex}{$\bullet$}}}

\newcommand{\btjnbrR}{{\mathrel{\rotatebox[origin=c]{90}
{$\bowtie$}}\kern-0.22em\raisebox{.9ex}{$\bullet$}
\kern-1.em\raisebox{-.8ex}{$\bullet$}}}
\newcommand{\btjnbrL}{{\mathrel{\rotatebox[origin=c]{90}
{$\bowtie$}}\kern-0.22em\raisebox{-.8ex}{$\bullet$}
\kern-1.em\raisebox{+.9ex}{$\bullet$}}}

\def\minusp{{\mp}}
\def\a{\alpha}
\def\b{\beta}
\def\c{\chi}
\def\d{\delta}
\def\e{\epsilon}
\def\ve{\varepsilon}
\def\f{\phi}
\def\g{\gamma}
\def\h{\eta}
\def\i{\iota}
\def\k{\kappa}
\def\l{\lambda}
\def\L{\Lambda}
\def\m{\mu}
\def\n{\nu}
\def\p{\pi}
\def\ps{\psi}
\def\r{\rho}
\def\s{\sigma}
\def\t{\tau}
\def\th{\theta}
\def\w{\omega}
\def\x{\xi}
\def\z{\zeta}
\def\D{\Delta}
\def\F{\Phi}
\def\P{\Pi}
\def\G{\Gamma}
\def\W{\Omega}
\def\Th{\Theta}
\def\pd{\partial}
\def\tq{\tilde{q}}
\def\tf{\tilde{f}}
\def\bz{\bar{z}}
\def\vf{\varphi}
\def\fin{f_{\infty}}
\def\ma{{\mathcal{A}}}
\def\mc{{\mathcal{C}}}
\def\md{{\mathcal{D}}}
\def\me{{\mathcal{E}}}
\def\mf{{\mathcal{F}}}
\def\mg{{\mathcal{G}}}
\def\mh{{\mathcal{H}}}
\def\mi{{\mathcal{I}}}
\def\mj{{\mathcal{J}}}
\def\mk{{\mathcal{K}}}
\def\ml{{\mathcal{L}}}
\def\mm{{\mathcal{M}}}
\def\mn{{\mathcal{N}}}
\def\mo{{\mathcal{O}}}
\def\mp{{\mathcal{P}}}
\def\mq{{\mathcal{Q}}}
\def\mr{{\mathcal{R}}}
\def\ms{{\mathcal{S}}}
\def\mt{{\mathcal{T}}}
\def\mw{{\mathcal{W}}}
\def\mz{{\mathcal{Z}}}
\def\fA{\mathfrak{A}}
\def\fB{\mathfrak{B}}
\def\fN{{\mathfrak{N}}}
\def\fQ{{\mathfrak{Q}}}
\def\ff{{\mathfrak{f}}}
\def\lan{\langle}
\def\ran{\rangle}
\def\pb{\textbf{P}}
\def\nb{\textbf{n}}
\def\xb{\textbf{x}}
\def\yb{\textbf{y}}
\def\Db{\textbf{D}}
\def\Kb{\textbf{K}}
\def\Pb{\textbf{P}}
\def\Mb{\textbf{M}}
\def\Lb{\textbf{L}}
\def\bh{\bar{h}}
\def\sb{\bar{s}}
\def\ib{{\mathbb{I}}}
\def\rb{\mathbb{R}}
\def\tell{\tilde{\ell}}
\def\tr{{\rm tr~}}
\newcommand{\eq}[1]{eq.\eqref{#1}}
\newcommand{\Eq}[1]{Eq.\eqref{#1}}
\def\tenp{\otimes}
\def\tens{\oplus}
\def\bt{\mathrel{\rotatebox[origin=c]{90}{$:\bowtie:$}}}

\newcommand{\tc}[2]{\textcolor{#1}{#2}}
\newcommand{\remark}[1]{\tc{red}{[#1]}}
\newcommand{\highlight}[1]{\tc{ForestGreen}{#1}}
\newcommand{\added}[1]{\tc{blue}{#1}}
\newcommand{\deleted}[1]{\tc{blue}{\sout{#1}}}
\newcommand{\replaced}[2]{\tc{blue}{\sout{#1}}\ \tc{blue}{{#2}}}

\def\Kagome{Kagom\'e}
\def\kagome{kagom\'e}
\def\hoct{\(h_{\mathrm{oct}}\)}
\def\mhoct{h_{\mathrm{oct}}}
\def\moct{\(\mathcal{M}_{\mathrm{oct}}\)}
\def\mmoct{\mathcal{M}_{\mathrm{oct}}}
\def\hxyz{\(h_{\mathrm{XYZ}}\)}
\def\mhxyz{h_{\mathrm{XYZ}}}
\def\mxyz{\(\mathcal{M}_{XYZ}\)}
\def\noncoplanar{non-coplanar}
\def\Octahedral{Octahedral}
\def\octahedral{Octahedral} \def\eGSTwelve{{$\approx\! -0.186221J_\chi$}}
\def\eGSThirtySix{{$\approx\! -0.172852J_\chi$}}
\begin{document}

\title{Chiral Broken Symmetry Descendants of the Kagom\'e Lattice Chiral Spin Liquid}
\begin{abstract}
{\bf 
The breaking of chiral and time-reversal symmetries provides a pathway to exotic quantum phenomena and topological phases. In particular, the breaking of chiral (mirror) symmetry in quantum materials has been shown to have important technological applications. Recent work has extensively explored the resulting emergence of chiral charge orders and chiral spin liquids on the kagom\'e lattice. 
Such chiral spin liquids are closely 
tied to bosonic fractional quantum Hall states and host anyonic quasiparticles;
however, their connection to nearby magnetically ordered 
states has remained a mystery. Here, we
show that two distinct \noncoplanar~magnetic orders with uniform spin chirality, the XYZ umbrella state and the 
Octahedral spin crystal,
emerge as competing orders in close proximity to the kagom\'e chiral spin liquid. Our results highlight the intimate
link between a many-body topologically ordered liquid and broken symmetry states with nontrivial real-space topology.}
\end{abstract}

\author{Anjishnu Bose}
\email{anjishnu.bose@mail.utoronto.ca}
\affiliation{Department of Physics, University of Toronto, 60 St. George Street, Toronto, ON, M5S 1A7 Canada}
\author{Arijit Haldar}
\email{arijit.haldar@utoronto.ca}
\affiliation{Department of Physics, University of Toronto, 60 St. George Street, Toronto, ON, M5S 1A7 Canada}
\author{Erik S. S{\o}rensen}
\email{sorensen@mcmaster.ca}
\affiliation{Department of Physics, McMaster University, 
1280 Main St. W., Hamilton ON L8S 4M1, Canada}
\author{Arun Paramekanti$^*$}
\email{arun.paramekanti@utoronto.ca}
\affiliation{Department of Physics, University of Toronto, 60 St. George Street, Toronto, ON, M5S 1A7 Canada}

\date{\today}
\maketitle

\vspace{-2mm}\section*{\hspace{-7mm} Introduction}\vspace{-2mm}

Quantum spin liquids (QSLs) are strongly entangled phases of 
quantum magnets which exhibit exotic quasiparticle excitations \cite{Balents_frustratedQSL_review_2010,Grover_EntanglementQSL_Review2013,Review_Savary_2016,review_Chamorro2021}. 
The classic work of Kalmeyer and Laughlin revealed a 
direct relation between a class of such QSLs,
with broken mirror and time-reversal symmetries,
and gapped fractional quantum Hall states of bosons with anyon excitations 
\cite{KalmeyerLaughlin}.
Important
progress was later made in identifying 
microscopic models on different lattices for which such
chiral spin liquids (CSLs) are exact 
\cite{SchroeterCSL2007,YaoKivelson_CSLNonAbelian_PRL2007,ThomaleCSL2009}
or numerically tractable 
\cite{Vidal2013,Bauer2014,YHe2014,Gong2015,Wietek_CSLKagome_PRB2015,YCHe_kagomeCSL_GaugedSPT_PRL2015,Wietek_CSL_Triangular_PRB2017,Moore_CSL_TriangularHubbard_DMRG_PRX2020,Hickey_TetrahedralCSL_PRL2016} ground states.
A valuable development
was the identification
of the Kalmeyer-Laughlin liquid 
in an $SU(2)$ invariant model with a simple
three-spin scalar 
chiral exchange coupling 
on the 
geometrically frustrated kagom\'e lattice 
\cite{Bauer2014,YHe2014,Gong2015,Wietek_CSLKagome_PRB2015},
a network of corner-sharing
triangles reminiscent of a Japanese woven basket 
\cite{kagomestory}.
While the nearest neighbor Kagom\'e lattice Heisenberg model 
has been argued to host a Dirac spin liquid
\cite{Hastings_DiracRVB_PRB2000,hermelePRL2007,Becca_kagomeVMCDirac_PRB2011,YHe_PRX_DMRG_DiracCone_2017,Iqbal_DiracVMC_PRB2021},
the inclusion of longer-range couplings has been shown to
realize CSLs \cite{Lhuillier_KagomeCSL_SchwingerBoson_PRL2012,YHe2014,Gong2015,Wietek_CSLKagome_PRB2015}
arising from spontaneous breaking of
mirror and time-reversal symmetries.
A variety of these competing phases have been
proposed to occur in materials such as Herbertsmithite \cite{YSLee_Herbertsmithite_PRB2007,Mendels_Herbertsmithite_Gapless_NatPhys2020}
and Zn-Barlowite \cite{Kagome_Barlowite_Nat2022}.
Optical driving \cite{Classen_kagome_Optical_Ncomm2017}, proximity to
Mott transitions \cite{Moore_CSL_TriangularHubbard_DMRG_PRX2020}, and twisted Moir\'e
crystals \cite{Zhang_MoireCSL_PRL2021} are potential experimental
routes to obtain CSLs, and even topological superconductors
upon doping 
\cite{HCJiang_CSLDoped_DMRG_PRL2020,YahuiZhang_DopedCSL_PRB2021}. More recently,
Rydberg atom quantum simulators have shown the promise
to access such topological spin liquids \cite{Lukin_Rydberg_kagomeLinks_Science2021}.

\begin{figure}
\centering
\includegraphics[width=\linewidth]{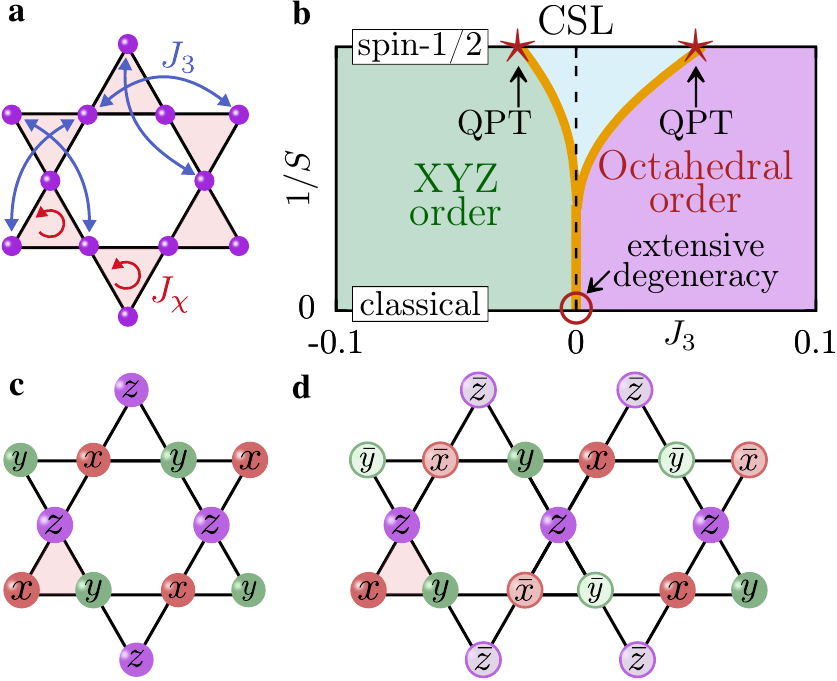}
    \caption[]
        {{\bf Kagom\'e lattice.} {\bf a} The chiral three-spin 
        interactions and further neighbor two-spin couplings
        in the model $H_{\rm spin}$.
       {\bf b} Proposed phase diagram of the model $H_{\rm spin}$
        as we tune $J_3$ and the strength of quantum fluctuations
        via $1/S$ where $S$ is the spin length. For spin
        $S\!=\!1/2$, the extensively degenerate 
        classical point evolves
        into an emergent chiral spin liquid, 
        bounded by \noncoplanar~magnetic orders. Light blue
        region indicates where spin-wave fluctuations can destabilize
        the \noncoplanar~orders hinting at a quantum spin liquid. 
        {\bf c} The XYZ umbrella magnetic ordering pattern indicating spins pointing 
        along $x,y,z$.
       {\bf d} The Octahedral magnetic ordering pattern with 
        quadrupled unit cell and spins pointing 
        along $(x,\bar{x})$, $(y,\bar{y})$, $(z,\bar{z})$
        where $\bar{x} \!\equiv - x$.}
        \label{fig:kagome oct}\label{fig:kagome}
\end{figure}

In parallel with the interest in such CSLs, there has been a great interest in chiral
broken symmetry states in geometrically frustrated systems, which can potentially
display nontrivial real-space topology. The most well known examples of these are 
skyrmion and meron crystals -- creating and manipulating such topological textures 
has important spintronics and information storage applications
\cite{Nagaosa_skyrmionreview_2013,Fert_Skyrmionreview_2017,Kurumaji_triangularskyrmionxtal_Science2019,Hirschberger_skyrmionbreathingkagome_NComm2019,Motome_skyrmionmeroncrystal_phase_Ncomm2021,SkyrmionWrite_Wang_NatComm2022}.
More recently, chiral density-wave orders have been reported 
in the metallic kagom\'e materials AV$_3$Sb$_5$ \cite{Ortiz_2019,Ortiz_2021,Hasan_ChiralCDW_kagome_AV3Sb5_NatMat2021,Luetkens_AV3Sb5_CDWTbreaking,Neupert_2022}, 
prompting a search for analogous chiral magnetic
orders in kagom\'e magnets such as FeGe \cite{PDai_FeGeMagnetic_2022}, and 
giving rise to the nascent field of ``chiraltronics''.

How are the topologically ordered states such as CSLs related to chiral broken symmetry states 
with nontrivial real-space topology? Historically, there was an attempt to relate
the fractional quantum Hall liquid to a melted Wigner 
crystal of electrons driven by multi-particle exchanges \cite{Kivelson_RingExch_FQH_PRB1987}. The analogous question in the field of QSLs is 
to ask how they arise from the melting of ``parent'' magnetically ordered states. For
instance, gapped $Z_2$ QSLs descend
from quantum melting of coplanar magnetic orders while 
preserving topological defects \cite{Chubukov_QPT_Noncollinear_NuclPhys1994}.
Here, we show that the kagom\'e lattice CSL
is in close proximity to two distinct 
symmetry breaking orders which
feature a uniform and nonzero scalar spin chirality, a 
nontrivial real-space topological
feature they partially share with skyrmion crystals \cite{Nagaosa_skyrmionreview_2013,Fert_Skyrmionreview_2017}.
The scalar chirality is a 
source of Berry fluxes, which may potentially 
transmute to background gauge
fluxes in an effective gauge theory description of the
spin-$1/2$ CSL 
\cite{WWZ_CSL_PRB1989,Fradkin_CSL_Deconfinement_PRL1991,YCHe_kagomeCSL_GaugedSPT_PRL2015,TaoLi_Tetrahedral_CSL_PRB2021}.
Our work links a many-body topologically ordered
state to the quantum melting of proximate chiral broken symmetry states 
with nontrivial real-space topology, and shows how both of these ultimately emerge 
from a highly degenerate manifold of classical chiral states.

\vspace{-2mm}\section*{\hspace{-7mm} Results}\vspace{-2mm}

\noindent{\bf Model Hamiltonian and Classical Orders} -- 
We consider the kagom\'e lattice model Hamiltonian
\begin{align}
\label{eq:spinham}
    H_{\rm spin} = -J_\chi\sum_{\bigtriangleup,\bigtriangledown}
    \vS_i\cdot \vS_j \times \vS_k 
    + J_3\sum_{\btjnbrR~~~\btjnbrL} \vS_i\cdot \vS_j.
\end{align}
Fig.~\ref{fig:kagome}{\bf a} shows
the chiral three-spin interaction $J_\chi$ 
acting on triangular plaquettes (with spins [$ijk$] 
ordered anticlockwise), and the $J_3$ Heisenberg 
term coupling farther spins on
 kagom\'e bow-ties. Without loss of generality, we fix $J_\chi\!=\!1$. 
Our proposed phase diagram for this
model is depicted in Fig.~\ref{fig:kagome}{\bf b}, as we tune $J_3$
and the spin length $S$
which controls the degree of quantum fluctuations. It 
prominently features two distinct 
chiral broken symmetry orders, and the
CSL in the spin-$1/2$ limit.

When $J_3\!=\!0$ in Eq.~\ref{eq:spinham}, 
minimizing the energy
amounts to maximizing the scalar spin chirality.
In the classical limit, where we treat spins as classical
unit vectors, this
implies that each triangle has spins which must form an
orthonormal triad, e.g., going anticlockwise around
a triangle, we may have spins pointing along $\{x,y,z\}$.
As shown in recent work 
on the kagom\'e lattice \cite{pitts2021order}, 
there can be many choices for how to place these triads 
on adjacent triangles, so this does not uniquely
determine the ground state; the number of classical 
ground states scales as $\Omega \! \sim \! 2^{N/3}$
where $N$ is the number of kagom\'e sites.

However, we see that 
any nonzero $J_3 \! < \! 0$ completely breaks this
degeneracy, selecting a unique ground
state (upto global rotations)
with XYZ order as shown in 
Fig.~\ref{fig:kagome}{\bf c}. This XYZ state is a 
specific member of the family of ${\mathbf Q} \! = \! 0$ 
``umbrella states'' which have the same unit cell as
the original kagom\'e lattice.
In the opposite limit, when $J_\chi\!=\!0$, the kagom\'e lattice
decouples into three rhombic sublattices, each of which 
individually supports ferromagnetic order 
driven by $J_3 \! < \! 0$. 
In this limit, introducing
an infinitesimal $J_\chi$ couples the three sublattices,
again leading to XYZ order. The depicted
XYZ state can thus
be shown to be the unique classical ground state 
of $H$ for any $J_3\!<\!0$
since it separately minimizes
each term in the Hamiltonian. A similar analysis indicates that
$J_3 \! > \! 0$ leads to antiferromagnetically coupled rhombic
sublattices. This selects \Octahedral~order, with a
$12$-site unit cell
and zero net magnetization, as the unique classical ground state.
Spins in the XYZ state subtend a solid angle $\pi/2$ over
elementary triangular plaquettes and trace out $-\pi$ over hexagons.
With Octahedral order, spins subtend a solid angle $\pi/2$ over
triangular plaquettes and trace out $+\pi$ over hexagons.
The XYZ and \Octahedral~states are `regular magnetic orders' \cite{Misguich_RegularMagneticOrders_PRB2011},
where lattice symmetries are only broken due to broken spin rotation
symmetries; restoring spin rotation symmetry via quantum fluctuations
is thus expected to result in symmetric quantum spin liquids.

\noindent{\bf Quantum Fluctuations} --
Leading order quantum fluctuations in spin models may be treated using linear spin wave theory (SWT) 
which is exact to ${\cal O}(1/S)$. 
To formally treat our model Hamiltonian within SWT, we rescale 
$J_\chi \!\to\! J_{\chi}/(2 S)$ in 
Eq.~\ref{eq:spinham}, which leaves the spin-$1/2$
model unchanged but allows the two-spin and three-spin terms
to compete in the $S \!\to\! \infty$ limit. We then treat the small
fluctuations around the XYZ and \Octahedral~orders by deriving and solving the bosonic 
Bogoliubov deGennes SWT Hamiltonian directly in
real-space (see Methods).
Using this
approach, we find that the \Octahedral~state spectrum admits three exact zero
modes, consistent with the expected number of Nambu-Goldstone 
modes of the fully broken spin rotational
symmetry, while the XYZ order
admits two zero modes, reflecting the modified
count of Nambu-Goldstone modes due to 
the nonzero net magnetization \cite{Watanabe_GoldstoneReview_AnnRev2020}. In addition to these zero
modes, there are spin wave modes at nonzero energy; when
$J_3 \!\to \! 0$, a macroscopically large number of 
these excitations descend in energy and
merge with the zero modes,
reflecting the extensive degeneracy of the
classical ground states \cite{pitts2021order}.

Dropping the exact zero modes on
finite size
systems, we have computed the SWT correction to 
the classical \Octahedral~and XYZ order parameters and
extrapolated the result to the thermodynamic limit; see
Supplementary Information (SI) \cite{SI} for details. We
respectively denote these as $M_{\pm}$
for $J_3 \!>\! 0$ and $J_3 \!<\! 0$. These
order parameters take the form
$M_\pm \!=\! S \!-\! \alpha_\pm(J_3)$, where the correction term
$\alpha_\pm$ depends on $J_3$ but is independent of $S$.
For small values of
$|J_3|$, these are well fit by the
expressions $\alpha_\pm (J_3) \! = \!c_{\pm} \ln(1/|J3|)$ 
where $c_+ \!\approx\! 0.068$ and $c_{-}\!\approx\! 0.053$; 
this logarithmic divergence as $J_3\!\to\!0^\pm$ is consistent with the absence
of long-range order at $J_3\!=\! 0$.

We identify the critical spin value $S_c$ where
these \noncoplanar~orders melt for a given $J_3$
using an analogue of the well-known Lindemann criterion for
melting of crystals. For the magnetic order to melt, we 
demand that $\alpha_\pm(J_3) \! >\! f S$, where $f$ is a 
constant.
This is equivalent to
demanding that the fluctuations exceed a sizeable
fraction of the classical ordered moment. Using this,
we obtain $1/S_c^{(\pm)}\!=\! (f/c_{\pm})/\ln(1/|J_3|)$.
For $f\!=\!0.4$, we find for spin $S\!=\!1/2$ that
this leads to loss of \Octahedral~order for 
$0 \!<\! J_3 \! \lesssim\! 0.05$ and a breakdown of the XYZ
order in the regime
$-0.02 \!\lesssim\! J_3 \!<\! 0$. In the $S\!=\!1/2$ model, 
we will see below that this (approximate) window  around $J_3\!=\!0$
gets replaced 
by the CSL. Plotting the melting curve for all $S$ leads to the 
phase boundaries marked in Fig.~\ref{fig:kagome}{\bf b}, which
reveals a spin liquid fan emanating from the
extensively degenerate classical chiral point.

\noindent{\bf Parton mean-field theory} --
To study the phase diagram of this model in the quantum
limit of $S\!=\!1/2$, we 
begin with a Schwinger fermion representation of the
spin $\vec S_i \!=\! f^\dagger_{i \alpha} \boldsymbol{\s}^\pdg_{\alpha\beta}
f^\pdg_{i \beta}/2$,
with an implicit sum on repeated (Greek) spin indices.
Previous DMRG and ED calculations \cite{Bauer2014} on the pure chiral
model with $J_3\!=\!0$ have shown that it supports a Kalmeyer-Laughlin
CSL ground 
state. Such a CSL is described at the mean-field level in terms of the fermionic ``$f$'' partons
as a topological band insulator with total Chern
number $C=2$. This topological insulator is obtained by 
filling half of the 
Chern bands formed by a uniform flux piercing elementary
triangular plaquettes of the kagom\'e lattice \cite{hermelePRL2007}.
To study the impact of $J_3$, we recast
the spin model in terms of partons
\bea
H_{\rm parton} &=& - \sum_{\la ij \ra} 
(t_{ij} f^\dg_{i\alpha} f^\pdg_{j\alpha} + t^*_{ij} f^\dg_{j\alpha} f^\pdg_{i\alpha}) \nonumber \\
&+& 
\frac{J_3}{4}~\sum_{\btjnbrL~~~\btjnbrR}~
f^\dg_{i\alpha} \vec\sigma_{\alpha\beta} f^\pdg_{i\beta} \cdot 
f^\dg_{j\mu} \vec\sigma_{\mu\nu} f^\pdg_{j\nu}.
\label{eq:Hparton}
\eea
Here, the first term is a kagom\'e Hofstadter model which
captures the mean field description of the
CSL at $J_3\!=\!0$ \cite{MarstonZeng_Kagome_DimerCSL_JAP1991,Hastings_DiracRVB_PRB2000}. The complex hoppings $t_{ij}$ are fixed to have equal 
magnitude $|t_{ij}|\!=\! t$ on all nearest-neighbor bonds,
and phases chosen such that the
partons experience $\pi/2$-flux around elementary 
triangular plaquettes and 
zero-flux around hexagonal plaquettes.
This supports Chern bands with 
total Chern number $C\!=\!2$ (counting both spin-$\upa$ and 
spin-$\dna$) at half-filling, providing 
the correct starting point for the low energy 
$U(1)_2$ Chern-Simons gauge theory description 
of the CSL \cite{WWZ_CSL_PRB1989}.
The mean-field spin gap in this insulator
is equal to its insulating band gap $\Delta_{mf} \approx \! 1.46 t$.
Matching this to
the ED result for the spin gap $\Delta\!\approx \! 0.05 J_\chi$ 
of the pure chiral model (see \cite{Bauer2014} and Fig.~\ref{fig:kagome ED}
below) fixes
$t\!=\!0.034 J_\chi$.
The second term in Eq.~\ref{eq:Hparton} 
is obtained by rewriting the $J_3$ spin 
interaction in Eq.~\ref{eq:spinham}
in terms of partons. This Hamiltonian
supplemented by a mean-field constraint 
$ \la f^\dg_{i \alpha} f^\pdg_{i \alpha} \ra \!=\! 1$ at each site.

\begin{figure}[t]
\centering
\includegraphics[width=0.48\textwidth]{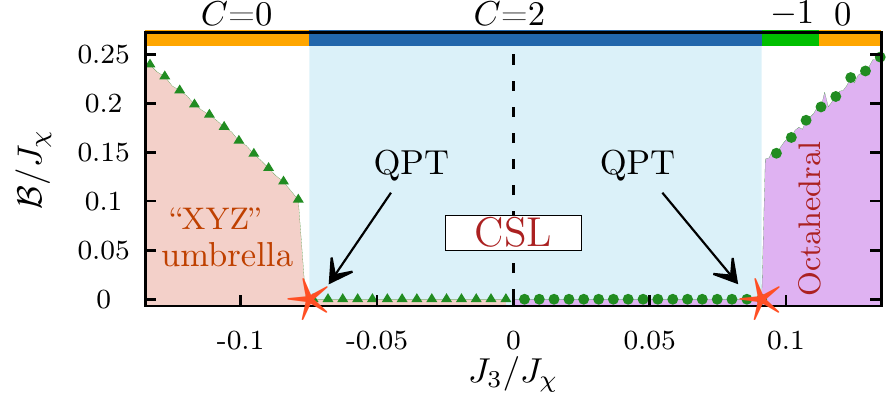}
    \caption
        {{\bf Parton theory phase diagram}.
        Mean field phase diagram of the $S=1/2$ parton 
       theory of Eq.~\ref{eq:Hparton}
        as we vary $J_3/J_\chi$. For $J_3 \!>\!0.092 J_\chi$, 
        we find a phase transition from the mean field CSL into the \Octahedral~state,
        while for $J_3 \!<\!-0.077 J_\chi$, we find an instability into ``XYZ'' umbrella
        order, a state with the same symmetries 
        as the XYZ state.
Top line depicts the total Chern number 
        of the half-filled parton bands in various phases as
        we tune $J_3/J_\chi$.}
\label{fig:parton}
\end{figure}

To examine the impact of $J_3$, we 
treat the four-fermion terms
using a spatially 
inhomogeneous and unbiased variational
mean-field theory on system sizes upto $108$ sites (see Methods).
For small $|J_3|$, 
the gapped Chern insulator is 
stable to four-fermion interactions. Beyond a critical coupling strength,
we find that the internal Weiss 
fields become nonzero, having a uniform strength ${\cal B}$ and directions
which are spatially modulated signifying magnetic symmetry breaking.
For $J_3\!>\!0.092 J_\chi$, we find that the converged
broken symmetry pattern shows a clear pattern of \Octahedral~order with a reconstructed $12$-site unit cell 
($2\! \times\! 2$ kagom\'e unit cell). For $J_3 \!<\!-0.077J_\chi$, 
the solution converges to ``XYZ'' order, i.e. an
umbrella state which shares all symmetries of the ${\mathbf Q}\!=\!0$ XYZ order, but
smoothly interpolates between the XYZ state and the ${\mathbf Q}\!=\! 0$ coplanar $120^\circ$ state.
We have also computed the total Chern number of the
occupied bands as we tune $J_3$ \cite{SI}. The resulting phase diagram
is shown in Fig.~\ref{fig:parton}.
Beyond mean-field theory, the broken symmetry
insulators
are expected to have 
trivial many-body topology.

\begin{figure}[t]
\centering
\includegraphics[scale=1]{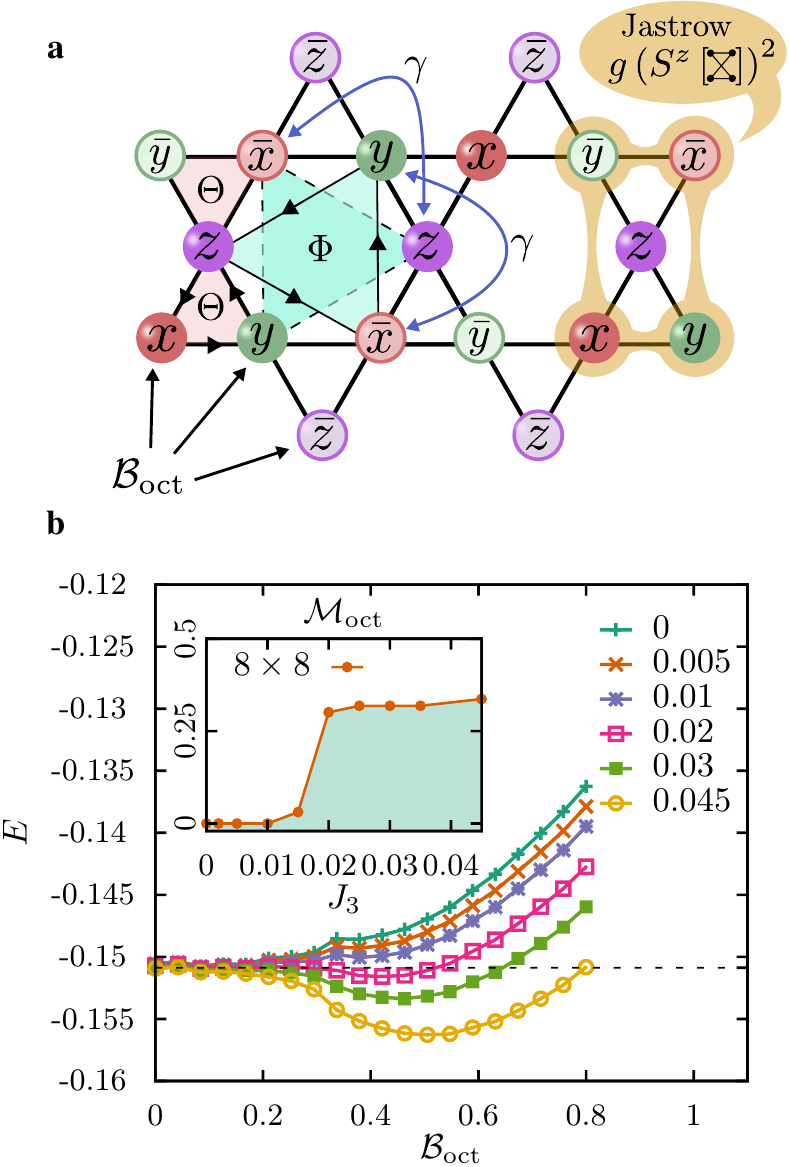}
    \caption[]
        {{\bf Gutzwiller wavefunction study.} {\bf a} Variational
        parameters used in our kagom\'e
        wavefunction ansatz include a second-neighbor hopping 
        of strength $\gamma$, fluxes $\Theta, \Phi$ 
        through elementary and large triangular plaquettes (shaded),
        a local Jastrow factor which suppresses the total
        $S^z$ on bow-ties (excluding the central site), and
        a Weiss field ${\cal B}_{\rm oct}$ which
        induces Octahedral order; field directions are depicted on the sites. {\bf b} Variational energy versus 
        ${\cal B}_{\rm oct}$ for an $8\times 8$
        lattice for different $J_3$, showing
        a stable CSL state for $J_3\!=\!0$ and an instability to
        Octahedral order for $J_3\! \gtrsim\! 0.02$. Inset
        shows the order parameter ${\cal M}_{\rm oct}$ 
        which becomes nonzero
        in the ordered phase.}
\label{fig:VMC}
\end{figure}

\noindent{\bf Gutzwiller projected wavefunctions} -- 
To go beyond parton mean-field theory, and strictly
implement the Gutzwiller projection constraint (i.e., exactly
one fermion per site), we next turn to
a Monte Carlo study of the projected parton wavefunction
\cite{ShengVMC2015}
to optimize its parameters and study its properties.
We consider a parton state $|\Psi_f\ra$ which
is obtained as the Slater determinant ground state of a 
variational Hamiltonian which includes complex nearest neighbor
hopping $e^{i\theta_{ij}}$
and next-neighbor hopping $\gamma e^{i\phi_{ij}}$ \cite{ShengVMC2015}, with
phases chosen to enclose uniform fluxes $\Theta,\Phi$ through 
elementary and large 
triangular plaquettes as in Fig.\ref{fig:VMC}{\bf a}. We
also include a Weiss field ${\mathbf b}_i$ via
$-\sum_i \vec b_i \cdot f^\dagger_{i \alpha} \boldsymbol{\s}^\pdg_{\alpha\beta}
f^\pdg_{i \beta}/2$ to account
for magnetic symmetry breaking orders,
limiting ourselves to \Octahedral~order ($J_3 \! > \! 0$) 
with zero net magnetization; the variational Weiss fields 
$\vec b_i$
are thus chosen to have an \Octahedral~pattern as indicated
in Fig.\ref{fig:VMC}{\bf a}, with a spatially 
uniform magnitude $|\vec b_i| \!=\! {\cal B}_{oct}$.
We explore the variational ansatz
\bea
|\Psi_G\ra = 
\prod_{\mathrel{\rotatebox[origin=c]{90}{$\bowtie$}}}
e^{-g \left(S^z[~{\btjstrw}~~]\right)^2}~P_G |\Psi_f\ra,
\eea
where $P_G$ denotes Gutzwiller projection to one electron per
site. The Gutzwiller
wavefunction $P_G |\Psi_f\ra$ 
is supplemented with a product
Jastrow correlation factor acting on every kagom\'e bow-tie, 
as shown in Fig.\ref{fig:VMC}{\bf a},
where $g$ is the strength of the Jastrow factor and
$S^z[~\btjstrw~~]$
denotes the total $S^z$ on the bow-tie excluding
the central site. The set of variational parameters explored
in our study are $\{\gamma,\Theta, \Phi, g, 
{\cal B}_{\rm oct}\}$ (see Methods).

For small $J_3/J_\chi$ of interest,
we find that we can get reasonable variational energies 
by fixing $\Theta\!=\!\pi/2$ and $\gamma\!=\!0.2$ in 
$H^{\rm var}_{\rm parton}$, and setting the Jastrow
strength to $g\!=\!0.045$. We then
vary $\Phi,{\cal B}_{\rm oct}$ 
to explore the variational
space for different values of $J_3$ (with $J_\chi\!=\!1$).
For the pure chiral model ($J_3\!=\!0$) 
our wavefunction on an $8\times 8$
kagom\'e lattice ($192$ spins) yields an
energy per site $\simeq -0.151(1) J_\chi$; this is somewhat
higher than ED ($\approx -0.1729 J_\chi,\ N=36$) and a previous iPEPS study
which yield $\approx \! -0.1715 J_\chi$.
Fig.~\ref{fig:VMC}{\bf b} shows the variational energy as 
a function of ${\cal B}_{\rm oct}$ for 
various values of $J_3$, where we have optimized
with respect to $\Phi$ at each point. For $J_3\!=\!0$,
we find that the
CSL is stable towards \Octahedral~magnetic ordering,
but with an apparent local metastable minimum 
at nonzero ${\cal B}_{\rm tot}$. With increasing $J_3$,
this metastable minimum rapidly comes down in energy, 
becoming the true minimum for $J_3/J_\chi \! \gtrsim\! 0.02$,
signalling a first-order transition into the 
\Octahedral~state. As shown in the inset
to Fig.~\ref{fig:VMC}{\bf b}, the 
\Octahedral~order 
${\cal M}_{\rm oct} \!=\! (1/N) \!\sum_i\! \vec m_i \cdot \hat{\vec b}_i$
jumps at this transition.
We recognize that a better CSL wavefunction at $J_3\!=\!0$ 
will have lower energy,
rendering the CSL more
stable and increasing the critical value of $J_3$
for the Octahedral instability.
We thus turn to a numerical exact diagonalization study to shed
further light on the $S\!=\!1/2$ phase diagram.

\noindent{\bf Exact Diagonalization Results} -- 
ED is a powerful unbiased tool to study frustrated kagom\'e quantum magnets 
\cite{Leung1993,Waldtmann_KagomeED_EPJB1998,Bauer2014,Lauchli2019}.
To corroborate our results from the preceding sections we have carried out ED calculations for the spin
Hamiltonian in Eq.~\ref{eq:spinham} on various
finite-size kagom\'e clusters, shown in Fig.~\ref{fig:kagome ED}{\bf f}, ranging in size from $N=12$ to $36$.
\begin{figure*}
\includegraphics[width=\linewidth]{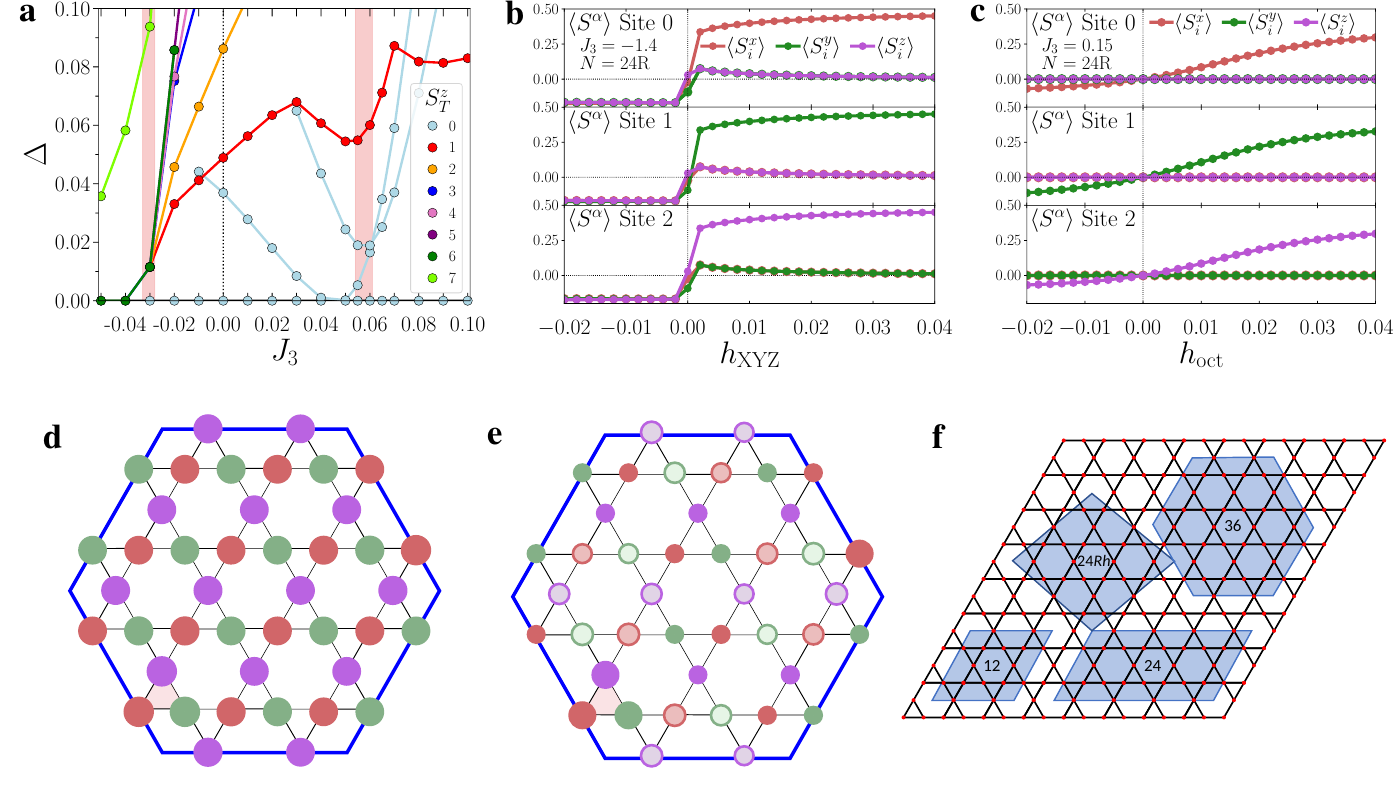}
    \caption{{\bf Exact diagonalization results}. 
    {\bf a} Gaps to the lowest lying $S^z_T=0,\ldots 7$ states for the 36-site cluster. 
    {\bf b} $\langle S^\alpha\rangle$ on adjacent sites taken anticlockwise on any given single up triangle 
    for the $24Rh$-site cluster in a {\it uniform} $h_{\rm XYZ}$ field with $J_3=-1.4$.
    {\bf c} $\langle S^\alpha\rangle$ on adjacent sites on a single triangle for the $24$-site cluster in a {\it uniform} $h_{\rm oct}$ field with $J_3=0.15$.
    {\bf d} The induced $\langle S^\alpha\rangle$ for the 36-site cluster with a Zeeman field
    $h_{\rm XYZ}$ applied to a {\it single} triangle (shaded) at $J_3=-0.5$ and
    {\bf e} with a $h_{\rm oct}$ applied to a {\it single} triangle (shaded) at $J_3=0.1$. The radius of the points are proportional to the overlap with the expected XYZ or \Octahedral~ordering direction at each site.
    {\bf f} The unit cells employed. Note that two different clusters with $N=24$, a rhombic labelled $24Rh$ and a rectangular labelled $24$. The rhombic $N=36$ unit cell is for convenience drawn in its equivalent hexagonal form in {\bf d, e, f}.
    }  
    \label{fig:kagome ED}
\end{figure*}
The largest clusters are studied using a fully parallelized Lanczos code that is most optimally used only with the total $S^z_T$ as a quantum number~\cite{LauchliED2011}. A full symmetry analysis can be performed on the smaller clusters (see SI~\cite{SI}). Our results for the spin gaps to the lowest lying states in each $S^z_T$ sector are shown in Fig.~\ref{fig:kagome ED}{\bf a} versus $J_3$. Two transitions are visible indicated by the shaded red regions. For $J_3\!\lesssim\! J^c_{\rm XYZ}\!\approx\! -0.03 J_\chi$ the ground-state transitions away from a singlet and the system becomes ferromagnetic, consistent with the appearance of the XYZ umbrella state. In the vicinity of 
$J_3\!=\!J^c_{\rm oct}\!\approx\! 0.06 J_\chi$ the spin gap appears to close signaling a {\it second order} transition to a different state. These values of $J^c_3$ compare favorably to the estimates obtained in our previous analysis. In the CSL-regime for 
$J^c_{\rm XYZ}\!<\!J_3\!<\!J^c_{\rm oct}$ our results are consistent with a finite spin-gap to the first $S=1$ state above two $S=0$ states.

To identify the magnetically ordered states adjoining the CSL phase we first apply a Zeeman field of the form $-\sum_\triangle h_{\rm XYZ}^\alpha S^\alpha$ on all up-triangles in the lattice, so that each site is counted once. On a single up-triangle with sites numbered (0,1,2), this
results in a contribution to the Hamiltonian $-h_{\rm XYZ}(S^x_0\!+\!S^y_1\!+\!S^z_2)$. 
Such a field term will induce the XYZ umbrella state at large $h_{\rm XYZ}$. For the three sites labelled 0,1,2 (anticlockwise) around a single triangle, the response $\langle S^\alpha\rangle$ versus $h_{\rm XYZ}$ is shown in 
Fig.~\ref{fig:kagome ED}{\bf b} at $J_3\!=\! -1.4J_\chi$ for the $24Rh$-cluster. Due to the degenerate ground-state at $J_3\!=\! -1.4J_\chi$ a discontinuous jump in all $\langle S^\alpha\rangle$ is observed at $h_{\rm XYZ}\!=\!0$ resulting in a divergent susceptibility with respect to the XYZ umbrella state. In a similar manner we can apply a field term of the form $-\sum_\triangle h_{\rm oct}^\alpha S^\alpha$ for $J_3>J^c_{\rm oct}$ with $h^\alpha_{\rm oct}$ now reflecting the \Octahedral~ordering shown in 
Fig.~\ref{fig:kagome oct}{\bf d}, with a pattern similar to the Weiss field ${\cal B}_{\rm oct}$ used in our variational study. The response of the system to such a field is shown in Fig.~\ref{fig:kagome ED}{\bf c} versus $h_{\rm oct}$ at $J_3\!=\!0.15J_\chi$ for the $24$-cluster. (The \Octahedral~ordering is not compatible with the $24Rh$-cluster). Because of the nonzero spin gap, the response is more gradual, but $\langle S^\alpha\rangle$ quickly reach values close to saturation even for small fields. For this value of $J_3$ we expect the spin gap to close with $N$. In the limit 
$h_{\rm oct}\!\to\! 0$ we can interpret $\partial \langle S^\alpha\rangle/\partial h_{\rm oct}$ as a susceptibility; 
we have verified that this susceptibility appears to diverge with 
$N$ \cite{SI}. On the other hand, if $h_{\rm XYZ}$ or $h_{\rm oct}$ is applied within the CSL, a first order transition 
to an ordered state is observed at a finite value of the field \cite{SI}.

To further study the magnetic ordering we apply a finite $h_{\rm XYZ}$ and $h_{\rm oct}$ to a {\it single} triangle (shown in shaded red in Fig.~\ref{fig:kagome ED}{\bf d,e} for the largest $N=36$ cluster and study the induced ordering at the other sites. This breaks most remaining symmetries, necessitating a diagonalization in the full $2^{36}$ dimensional Hilbert space. The results are shown in Fig.~\ref{fig:kagome ED}{\bf d,e} for $h_{\rm XYZ}, h_{\rm oct}=0.4$ at $J_3\!=\!-0.5$ and $J_3\!=\!0.15$, respectively. The observed patterns are clearly consistent with the XYZ umbrella and \Octahedral~ordering with only limited decrease in the overlap as one moves away from the triangle where the field is applied (shaded red). We calculate the induced ${\cal M}_{\rm oct}\!=\! 0.311$ at $J_3\!=\!0.15J_\chi$,
in good agreement with the Gutzwiller wavefunction result, and ${\cal M}_{\rm XYZ}\!=\! 0.443$
at $J_3\!=\!-0.5J_\chi$.
Our ED results
unequivocally point to the presence of XYZ and \Octahedral~orders in close proximity to the CSL.

\vspace{-2mm}\section*{\hspace{-7mm} Discussion}\vspace{-2mm}
In this work, we have 
used spin-wave theory, ED, 
and Gutzwiller wavefunctions, to
uncover two chiral magnetic orders -- 
XYZ order and \Octahedral~order -- 
near the gapped CSL on the 
kagom\'e lattice, which are accessed by tuning a small Heisenberg interaction across the bow-ties.
Our proposed global phase diagram, as we vary spin $S$, 
hints at the possibility of unusual QSLs in
the chiral model for higher spin, including spin-$1$ magnets, opening up a promising research direction.
Previous ED and DMRG calculations have found CSLs and 
tetrahedral spin crystals on
triangular and honeycomb lattices 
\cite{Vidal2013,Hickey_TetrahedralCSL_PRL2016,Wietek_CSL_Triangular_PRB2017,Hickey_Tetrahedral_PRB2017},
and complex \noncoplanar~orders
in kagom\'e lattices with staggered chiral terms which hosts a
gapless CSL \cite{Pereira_GaplessCSL_VMC2021}.
Our work unveils distinct \noncoplanar~orders on the kagom\'e lattice, and
points to a universal
connection between many-body 
topological order in the gapped CSL and real-space topology 
encoded in Berry fluxes of the \noncoplanar~broken symmetries. Further research is needed 
to establish such a connection within a field theoretic framework.
It would be valuable
to extend our work to explore competing orders in
models which spontaneously break these symmetries \cite{YHe2014,Gong2015}, and study
the impact of charge doping \cite{HCJiang_CSLDoped_DMRG_PRL2020}.
Finally, our work lends impetus to extend the exploration of 
kagome skyrmion materials \cite{Hirschberger_skyrmionbreathingkagome_NComm2019} 
to the quantum regime to study the
melting of skyrmion crystals as a route to CSLs.

\noindent {\bf Methods \label{sec:methods}}

\noindent{\it Spin wave theory}
To study quantum fluctuations around the XYZ and \Octahedral~orders, 
we first perform a local spin rotation $R_{j}$
to align all spins along a global \(z-\)axis, 
\(\Tilde{\boldsymbol{S}}_{n, j}=R_{j}\cdot\boldsymbol{S}_{n, j}\), where \(n\) refers to the 
magnetic unit cell, and \(j\) represents the sub-lattice \cite{LinearSW2015}. 
Expanding the Hamiltonian using Holstein-Primakoff 
bosons via
$\tilde{S}_{n, j}^{+} = \sqrt{2 S}~b_{n, j},
\tilde{S}_{n, j}^{-} = \sqrt{2 S}~b_{n, j}^{\dagger},
\tilde{S}_{n, j}^{z} =  S-b_{n,j}^{\dagger}b_{n, j}$,
we keep terms upto quadratic order in bosons. 
Diagonalizing the resulting Bogoliubov deGennes 
Hamiltonian \cite{Colpa1978,LinearSW2015} (see SI for details),
we calculate the order parameter correction $\alpha\equiv\sum_{n,j} \langle b^{\dagger}_{n,j}
b_{n,j} \rangle/N$.

\noindent{\it Parton mean-field theory} The mean-field 
calculation for Eq.(\ref{eq:Hparton}) assumes a uniform flux pattern of \([\pi/2, \pi/2, 0]\) through the up, down-triangles, and the hexagons of the kagom\'e lattice \cite{MarstonZeng_Kagome_DimerCSL_JAP1991,Hastings_DiracRVB_PRB2000}. The trial Hamiltonian consists of the same nearest-neighbour hopping as \(\mh_{\rm parton}\) while quartic fermion interactions are replaced by complex bow-tie hoppings and independent Zeeman fields 
on every site. We minimize \(\expval{\mh_{\rm parton}}\) in the ground state of the trial Hamiltonian, with respect to this large set of mean-field parameters for system sizes upto $108$ 
kagom\'e sites for various \(J_3\). This leads to the spontaneous
magnetically ordered states shown in the phase diagram.
The corresponding total Chern numbers at half-filling bands \cite{Fukui:2005wr} are also calculated in the converged solution,
and  shown in Fig.\ref{fig:parton}. The full Chern number phase diagram varying both flux and the Weiss 
field strength, and details of calculating \(\expval{\mh_{\rm parton}}\) are given in the SI \cite{SI}.

\noindent{\it Variational Monte Carlo study}
We use the Metropolis algorithm to stochastically sample spin configurations (typically $\sim 5 \times 10^4$)
in the $S_z$ basis in our variational
wavefunction $|\Psi_G\rangle$ (which is a
Slater determinant multiplied by the Jastrow prefactor)
in order to calculate the expectation value of the energy.
For \noncoplanar~states, it is convenient to interpret
spin-$\uparrow$ and spin-$\downarrow$ as an additional 
layer coordinate, and the in-plane components of the
variational Weiss fields as inter-layer hoppings.
To test our optimized chiral wavefunction against previously
reported results \cite{ShengVMC2015}, we generalized the spin
model $H_{\rm spin}$ to incorporate a nearest-neighbor 
Heisenberg exchange term with strength $J_1$, and
explored different values of $(J_1,J_3,J_\chi)$.
(i) For $J_3\!=\!0$ and $J_\chi\!=\!0.15 J_1$,
our optimal wavefunction 
yields an energy per
site $\approx \!- 0.445(1) J_1$, quite close
to a previous careful VMC study of the model \cite{ShengVMC2015}
which found  $\approx\! -0.450 J_1$.
(ii) On the $12$-site kagom\'e cluster, 
ED for the pure chiral model (i.e., $J_3\!=\!J_1\!=\!0$) 
yields a singlet
ground state, with energy per spin \eGSTwelve{}, while
our $12$-site spin liquid wavefunction yields 
$\approx\! -0.176(1) J_\chi$ per spin. For comparison, ED on the 36-site cluster at $J_3=0$ yields \eGSThirtySix{}.

\noindent{\it Exact Diagonalization} Numerical exact diagonalization (ED) were performed using a fully parallelized Lanczos code using an on-the-fly calculation of the action of the Hamiltonian matrix. Clusters used in the calculations
are shown in Fig.~\ref{fig:kagome ED}{\bf f}. The full symmetry analysis of the spectrum was done on a 
smaller, \(12-\)site system, and the $C_6$ rotation eigenvalues of the lowest two singlets
was found to be consistent with earlier studies \cite{Vidal2013}. Further details are presented in the Supplementary Information (SI) \cite{SI}.

\vspace{2mm}

\noindent {\bf Supplementary Information} is available in the online version of the paper. \\[-0.5mm]

\noindent {\bf Acknowledgements}

\noindent This work was supported by the Natural Sciences and Engineering Research Council of Canada.
This research was enabled in part by support provided by Sharcnet (\href{http://www.sharcnet.ca}{www.sharcnet.ca}) 
and Compute Canada (\href{http://www.computecanada.ca}{www.computecanada.ca}).

\noindent {\bf Author Contributions}

\noindent Exact diagonalization calculations were performed by E.S.S and A.B.
Spin wave and parton mean field calculations were carried out by
A.B. and A.H. The Gutzwiller Monte Carlo simulations were done by
A.P. and A.H.
A.P. planned and supervised the project.
All authors contributed to the writing of the manuscript. \\[-0.5mm]

\noindent {\bf Author Information}

\noindent The authors declare no competing financial interests. Correspondence should be addressed to A.P. (\href{mailto:arun.paramekanti@utoronto.ca}{arun.paramekanti@utoronto.ca}). \\[-0.5mm]

\noindent{\bf Data availability}

\noindent The data that support the findings of this study are available from the corresponding author upon reasonable request
and will later be made available on github.

\noindent{\bf Code availability}

\noindent The computer codes used to generate the data used in this study are available from the corresponding author upon reasonable request.
\ifdefined\makeSM{}
	\let\oldaddcontentsline\addcontentsline \renewcommand{\addcontentsline}[3]{}\bibliography{ref}
	\let\addcontentsline\oldaddcontentsline \else
	\bibliography{ref}
\fi
\ifdefined\makeSM{}
\clearpage
\newpage
\appendix
 \renewcommand{\appendixname}{}
 \renewcommand{\thesection}{{S\arabic{section}}}
 \renewcommand{\theequation}{\thesection.\arabic{equation}}
\renewcommand{\thefigure}{S\arabic{figure}}
 \setcounter{page}{1}
 \setcounter{figure}{0}
 \widetext

\baselineskip=15pt

\ifdefined\makeSM{}
\else
\documentclass[nofootinbib,preprint,aps,prl,superscriptaddress,onecolumn,longbibliography,draft=false, notitlepage]{revtex4-1}
\usepackage{color}
\usepackage{mathtools}
\usepackage{bm}        \usepackage{amssymb}   \usepackage{amsmath}
\usepackage{wasysym}
\usepackage[colorlinks=true,citecolor=blue,linkcolor=magenta]{hyperref}
\usepackage{graphicx}
\usepackage{braket}
\usepackage{natbib}
\usepackage [latin1]{inputenc}
\usepackage{siunitx}
\usepackage{blkarray, bigstrut}
\usepackage[dvipsnames]{xcolor}

\setcounter{secnumdepth}{2}
\setcounter{tocdepth}{2}

\usepackage{physics}
\graphicspath{{./Figures/}}

\usepackage{tikz}
\newcommand{\foo}[1]{\begin{tikzpicture}[#1]\draw (0,0) -- (1ex,1ex);\draw (0.5ex,0) -- (1.3ex,0.8ex);\end{tikzpicture}}

\newcommand{\bea}{\begin{eqnarray}}
\newcommand{\eea}{\end{eqnarray}}
\newcommand{\nn}{\langle ij \rangle}
\newcommand{\nnn}{\langle\langle ij \rangle\rangle}
\newcommand{\bsigma}{\boldsymbol{\sigma}}
\newcommand{\bS}{\mathbf{S}}
\newcommand{\bb}{\mathbf{b}}
\newcommand{\br}{\mathbf{r}}
\newcommand{\hatz}{\hat{z}}
\newcommand{\ba}{\mathbf{a}}
\newcommand{\bs}{\mathbf{s}}
\newcommand{\bp}{\mathbf{p}}
\newcommand{\bn}{\mathbf{n}}
\newcommand{\bG}{\mathbf{G}}
\newcommand{\bOm}{\mathbf{\Omega}}

\newcommand{\mb}{\mathfrak{b}}
\newcommand\bbone{\ensuremath{\mathbbm{1}}}
\newcommand{\ul}{\underline}
\newcommand{\vl}{v_{_L}}
\newcommand{\vc}{\mathbf}
\newcommand{\be}{\begin{equation}}
\newcommand{\ee}{\end{equation}}
\newcommand{\bk}{{{\bf{k}}}}
\newcommand{\bK}{{{\bf{K}}}}
\newcommand{\cE}{{{\cal E}}}
\newcommand{\bQ}{{{\bf{Q}}}}
\newcommand{\bg}{{{\bf{g}}}}
\newcommand{\hbr}{{\hat{\bf{r}}}}
\newcommand{\bR}{{{\bf{R}}}}
\newcommand{\bq}{{\bf{q}}}
\newcommand{\hx}{{\hat{x}}}
\newcommand{\hy}{{\hat{y}}}
\newcommand{\ha}{{\hat{a}}}
\newcommand{\hb}{{\hat{b}}}
\newcommand{\hc}{{\hat{c}}}
\newcommand{\hd}{{\hat{\delta}}}
\newcommand{\beal}{\begin{align}}
\newcommand{\eeal}{\end{align}}
\newcommand{\ra}{\rangle}
\newcommand{\la}{\langle}
\renewcommand{\tt}{{\tilde{t}}}
\newcommand{\upa}{\uparrow}
\newcommand{\dna}{\downarrow}
\newcommand{\vS}{\vec{S}}
\newcommand{\dg}{{\dagger}}
\newcommand{\pdg}{{\phantom\dagger}}
\newcommand{\tphi}{{\tilde\phi}}
\newcommand{\cf}{{\cal F}}
\newcommand{\ca}{{\cal A}}
\renewcommand{\ni}{\noindent}
\def\l{\ell}
\newcommand{\trs}{\mathcal{T}}
\newcommand{\bdel}{\boldsymbol{\delta}}
\renewcommand{\bm}{\mathbf{m}}

\def\Jt{\mathbf{J}}
\def\S{\tilde{\mathbf{S}}}
\def\Sc{\tilde{S}}
\def\h{\mathbf{h}}

\renewcommand{\vec}[1]{\mathbf{#1}}
\def\vS{\vec{S}}

\newcommand{\btjstrw}{\mathrel{{\rotatebox[origin=c]{90}
{$\bowtie$}}\kern-0.18em\raisebox{-.95ex}{$\bullet$}
\kern-0.5em\raisebox{.97ex}{$\bullet$}
\kern-1.12em\raisebox{.97ex}{$\bullet$}
\kern-0.52em\raisebox{-.95ex}{$\bullet$}}}

\newcommand{\btjnbrR}{{\mathrel{\rotatebox[origin=c]{90}
{$\bowtie$}}\kern-0.22em\raisebox{.9ex}{$\bullet$}
\kern-1.em\raisebox{-.8ex}{$\bullet$}}}
\newcommand{\btjnbrL}{{\mathrel{\rotatebox[origin=c]{90}
{$\bowtie$}}\kern-0.22em\raisebox{-.8ex}{$\bullet$}
\kern-1.em\raisebox{+.9ex}{$\bullet$}}}

\def\minusp{{\mp}}
\def\a{\alpha}
\def\b{\beta}
\def\c{\chi}
\def\d{\delta}
\def\e{\epsilon}
\def\ve{\varepsilon}
\def\f{\phi}
\def\g{\gamma}
\def\h{\eta}
\def\i{\iota}
\def\k{\kappa}
\def\l{\lambda}
\def\L{\Lambda}
\def\m{\mu}
\def\n{\nu}
\def\p{\pi}
\def\ps{\psi}
\def\r{\rho}
\def\s{\sigma}
\def\t{\tau}
\def\th{\theta}
\def\w{\omega}
\def\x{\xi}
\def\z{\zeta}
\def\D{\Delta}
\def\F{\Phi}
\def\P{\Pi}
\def\G{\Gamma}
\def\W{\Omega}
\def\Th{\Theta}
\def\pd{\partial}
\def\tq{\tilde{q}}
\def\tf{\tilde{f}}
\def\bz{\bar{z}}
\def\vf{\varphi}
\def\fin{f_{\infty}}
\def\ma{{\mathcal{A}}}
\def\mc{{\mathcal{C}}}
\def\md{{\mathcal{D}}}
\def\me{{\mathcal{E}}}
\def\mf{{\mathcal{F}}}
\def\mg{{\mathcal{G}}}
\def\mh{{\mathcal{H}}}
\def\mi{{\mathcal{I}}}
\def\mj{{\mathcal{J}}}
\def\mk{{\mathcal{K}}}
\def\ml{{\mathcal{L}}}
\def\mm{{\mathcal{M}}}
\def\mn{{\mathcal{N}}}
\def\mo{{\mathcal{O}}}
\def\mp{{\mathcal{P}}}
\def\mq{{\mathcal{Q}}}
\def\mr{{\mathcal{R}}}
\def\ms{{\mathcal{S}}}
\def\mt{{\mathcal{T}}}
\def\mw{{\mathcal{W}}}
\def\mz{{\mathcal{Z}}}
\def\fA{\mathfrak{A}}
\def\fB{\mathfrak{B}}
\def\fN{{\mathfrak{N}}}
\def\fQ{{\mathfrak{Q}}}
\def\ff{{\mathfrak{f}}}
\def\lan{\langle}
\def\ran{\rangle}
\def\pb{\textbf{P}}
\def\nb{\textbf{n}}
\def\xb{\textbf{x}}
\def\yb{\textbf{y}}
\def\Db{\textbf{D}}
\def\Kb{\textbf{K}}
\def\Pb{\textbf{P}}
\def\Mb{\textbf{M}}
\def\Lb{\textbf{L}}
\def\bh{\bar{h}}
\def\sb{\bar{s}}
\def\ib{{\mathbb{I}}}
\def\rb{\mathbb{R}}
\def\tell{\tilde{\ell}}
\def\tr{{\rm tr~}}
\newcommand{\eq}[1]{eq.\eqref{#1}}
\newcommand{\Eq}[1]{Eq.\eqref{#1}}
\def\tenp{\otimes}
\def\tens{\oplus}
\def\bt{\mathrel{\rotatebox[origin=c]{90}{$:\bowtie:$}}}

\newcommand{\tc}[2]{\textcolor{#1}{#2}}
\newcommand{\remark}[1]{\tc{red}{[#1]}}
\newcommand{\highlight}[1]{\tc{ForestGreen}{#1}}
\newcommand{\added}[1]{\tc{blue}{#1}}
\newcommand{\deleted}[1]{\tc{blue}{\sout{#1}}}
\newcommand{\replaced}[2]{\tc{blue}{\sout{#1}}\ \tc{blue}{{#2}}}

\def\Kagome{Kagom\'e}
\def\kagome{kagom\'e}
\def\hoct{\(h_{\mathrm{oct}}\)}
\def\mhoct{h_{\mathrm{oct}}}
\def\moct{\(\mathcal{M}_{\mathrm{oct}}\)}
\def\mmoct{\mathcal{M}_{\mathrm{oct}}}
\def\hxyz{\(h_{\mathrm{XYZ}}\)}
\def\mhxyz{h_{\mathrm{XYZ}}}
\def\mxyz{\(\mathcal{M}_{XYZ}\)}
\def\noncoplanar{non-coplanar}
\def\Octahedral{Octahedral}
\def\octahedral{Octahedral} 
\begin{document}
\fi

\ifdefined\makeSM{}
\centerline{\Large \bf Supplementary Information:}
\centerline{\Large \bf Chiral Broken Symmetry Descendants of the Kagom\'e Lattice}
\centerline{\Large \bf Chiral Spin Liquid}
\centerline{}
\centerline{Anjishnu Bose$^1$, Arijit Haldar$^1$, Erik S. S{\o}rensen$^2$, and Arun Paramekanti$^1$}
\centerline{}
\centerline{${}^1$\emph{Department of Physics, University of Toronto, 60 St. George Street, Toronto, ON, M5S 1A7 Canada}}
\centerline{${}^2$\emph{Department of Physics, McMaster University, 
280 Main St. W., Hamilton ON L8S 4M1, Canada}}

\else
\title{Supplementary Information: \\Chiral Broken Symmetry Descendants of the Kagom\'e Lattice Chiral Spin Liquid}
\author{Anjishnu Bose}
\affiliation{Department of Physics, University of Toronto, 60 St. George Street, Toronto, ON, M5S 1A7 Canada}
\author{Arijit Haldar}
\affiliation{Department of Physics, University of Toronto, 60 St. George Street, Toronto, ON, M5S 1A7 Canada}
\author{Erik S. S{\o}rensen}
\affiliation{Department of Physics, McMaster University, 
1280 Main St. W., Hamilton ON L8S 4M1, Canada}
\author{Arun Paramekanti}
\affiliation{Department of Physics, University of Toronto, 60 St. George Street, Toronto, ON, M5S 1A7 Canada}

\date{\today}
\maketitle
\fi
\tableofcontents

\section{Spin Wave theory}
To find the linear spin-wave dispersion, we first have to rotate each spin within a unit cell to the $\hat{z}$ direction. We can define three useful unit vectors quantities from the local rotation matrix \(R_j\) which rotates the ordered spin vector $\vec S_j \to \hat{z}$ \cite{LinearSW2015}, namely
\begin{equation}
    \label{Rj vecs}
    \begin{gathered}
        u_j^{\a} = R_{j}^{\a 1}+\i R_{j}^{\a 2}\,,\:\:\:\bar{\boldsymbol{u}}_j = \boldsymbol{u}_{j}^{*}\,,\:\:\:        v_{j}^{\a} = R_{j}^{\a 3}\,.
    \end{gathered}
\end{equation}
Here $j$ refers to the basis site within the magnetic unit cell, and $\alpha$ labels vector components, and superscripts $1,2,3$ refer to columns of the rotation matrix.
In terms of these, we can write the original spin operators as
\begin{equation}
    \label{ferro rot vec}
    S_{n,j}^{\a} = \frac{1}{2}\left(u_j^{\a}\tilde{S}_{n,j}^{-}+\bar{u}_j^{\a}\tilde{S}_{n,j}^{+} \right)+v_{j}^{\a} \tilde{S}_{n,j}^{z}
\end{equation}
Finally, we use the Holstein-Primakoff transformation as given in the main text as
\begin{equation}
    \label{HP transform}
    \begin{gathered}
        \tilde{S}_{n, j}^{+} = \sqrt{2 S}\: b_{n, j}\,,\\
        \tilde{S}_{n, j}^{-} = \sqrt{2 S}\: b_{n, j}^{\dagger}\,,\\
        \tilde{S}_{n, j}^{z} =  S-b_{n,j}^{\dagger}b_{n, j}\,.\\
    \end{gathered}
\end{equation}
and end up with
\begin{equation}
    \label{HP transform spin}
    S_{n,j}^{\a} = \sqrt{\frac{S}{2}}\left(\bar{u}_{j}^{\a} b_{n, j}+u_{j}^{\a} b_{n,j}^{\dagger}\right)+v_{j}^{\a} \left(S-b_{n,j}^{\dagger}b_{n, j}\right)
\end{equation}

\subsection{Chiral Hamiltonian}
We start with the pure chiral Hamiltonian on the \kagome~lattice as
\begin{equation}
    \label{chiral ham}
    \mh_{\c} = - \sum_{m,n,p}\sum_{i,j,k}J^{ijk}_{mnp}\:\boldsymbol{S}_{m,i}\cdot\left(\boldsymbol{S}_{n,j}\times\boldsymbol{S}_{p,k}\right)\,,
\end{equation}
where \(m,n,p\) mark the magnetic unit cell (quadrupled unit cell as compared to the normal \kagome), and \(i,j,k\in\{0,1,2,...,11\}\) for the \octahedral~order, whereas \(i,j,k\in\{0,1,2\}\) for the XYZ order,  mark the sublattices. \(J^{ijk}_{mnp}\) is chosen such that each up and down triangle is summed over once; this coupling constant will
be chosen to be $J_\chi$. After substituting back \eqref{HP transform spin} into \eqref{chiral ham}, and only keeping terms upto quadratic order (also ignoring linear terms since their expectation vanishes), we end up with
\begin{equation}
    \label{hamil boson }
    \begin{split}
        \mh_{\c} =  - \sum_{m,n,p}\sum_{i,j,k}J^{ijk}_{mnp}\: \bigg\{\frac{S^2}{2}\bigg[\bigg( & b_{n,j}b_{p,k}\left(\boldsymbol{v}_i\cdot\bar{\boldsymbol{u}}_{j}\times\bar{\boldsymbol{u}}_k\right)
        +  b_{n,j}b_{p,k}^{\dagger}\left(\boldsymbol{v}_i\cdot\bar{\boldsymbol{u}}_{j}\times\boldsymbol{u}_k\right)\\
        + & b_{n,j}^{\dagger}b_{p,k}\left(\boldsymbol{v}_i\cdot\boldsymbol{u}_{j}\times\bar{\boldsymbol{u}}_k\right)
        +  b_{n,j}^{\dagger}b_{p,k}^{\dagger}\left(\boldsymbol{v}_i\cdot\boldsymbol{u}_{j}\times\boldsymbol{u}_k\right)\bigg)\\
        + & \{\text{cyclic permutations}\}\bigg]\\
         +\left(\boldsymbol{v}_i\cdot \boldsymbol{v}_j\times\boldsymbol{v}_k\right) \bigg[& S^3- S^2 \left(  b_{m,i}^{\dagger} b_{m, i}+ b_{n,j}^{\dagger} b_{n,j}+ b_{p,k}^{\dagger} b_{p, k}\right)\bigg]\bigg\}
    \end{split}
\end{equation}

\subsection{Bow-tie Heisenberg Hamiltonian}
The bow-tie Heisenberg Hamiltonian looks like
\begin{equation}
    \label{bowtie ham}
    \mh_{bt} = \frac{1}{2}\sum_{m,n}\sum_{i,j} J_{m,n}^{i,j}\: \boldsymbol{S}_{m,i}\cdot \boldsymbol{S}_{n,j}\,.
\end{equation}
where the factor of \(1/2\) is because we will be using the symmetric form of \(J_{m,n}^{i,j} = J_{n,m}^{j,i}\), and \(J\) is chosen such that we sum over each bow-tie pair once, and we will fix this coupling constant to be $J_3$.
Again, repeating the steps as before, dropping terms higher order than quadratic in the boson operators, and also ignoring the linear terms, we end up with
\begin{equation}
    \label{bowtie Hamiltonian boson}
    \begin{split}
    \mh_{bt}  = \sum_{m,n}\sum_{i,j} \frac{J_{m,n}^{i,j}}{2}\:\bigg\{\frac{S}{2}\bigg[& b_{m,i} b_{n,j} (\bar{\boldsymbol{u}}_i\cdot\bar{\boldsymbol{u}}_j) + b_{m,i} b_{n,j}^{\dagger} (\bar{\boldsymbol{u}}_i\cdot\boldsymbol{u}_j)
    +  b_{m,i}^{\dagger} b_{n,j} (\boldsymbol{u}_i\cdot\bar{\boldsymbol{u}}_j)+b_{m,i}^{\dagger} b_{n,j}^{\dagger} (\boldsymbol{u}_i\cdot\boldsymbol{u}_j)\bigg]\\
     + (\boldsymbol{v}_i\cdot\boldsymbol{v}_j)\bigg[& S^2- S \bigg( b_{m,i}^{\dagger}b_{m,i}+ b_{n,j}^{\dagger}b_{n,j}\bigg)\bigg]\bigg\}\,.
    \end{split}
\end{equation}

\subsection{Bogoliubov deGennes (BdG) Hamiltonian}
Combining the two as \(\mh=1/(2S)\cdot\mh_{\c}+ \mh_{bt}\), we end up with a BdG Hamiltonian. To diagonalize it, we first have to ensure that the Hamiltonian is positive definite. In our case, it is actually positive semi-definite due to the existence of Goldstone modes. The number of Goldstone modes depends on the ordering \cite{Watanabe_GoldstoneReview_AnnRev2020} which can be seen in the formula as
\begin{equation}
    \label{goldstone count}
    n_{GM} = n_{BG}-\frac{1}{2}{\mathrm{Rank}}(\r)\,,
\end{equation}
where \(n_{GM}\) is the number of Goldstone modes, \(n_{BG}\) is the number of broken generators. Furthermore, 
\begin{equation}
    \label{rho def}
    \r^{ab} = \frac{1}{V}f^{abc}\expval{Q^c}\,,
\end{equation}
where \(V\) is the volume of the system, \(f^{abc}\) are the structure constants of the symmetry group being broken, and \(Q^c\) are the generators of the global symmetry. For a model on a lattice, \(V=N\), the number of sites. Since the symmetry group being broken in our case is \(SU(2)\) spin-rotation symmetry, we also have that \(f^{abc}=\epsilon^{abc}\), and \(Q^c = \sum_{i=1}^{N}S_i^c\).

On the ferromagnetic side with the XYZ ordering,
\(\expval{Q^c} = S \frac{N}{3}\), for each of $c=x,y,z$ because we will have \(N/3\) spin\(-S\) pointing along the three orthogonal directions. This gives \({\mathrm{Rank}}(\r)=2\). However, since XYZ ordering completely breaks all spin-rotational symmetries, \(n_{BG}=3\). Hence using Eq.~\ref{goldstone count}, we find that \(n_{GM}^{\mathrm{XYZ}}=2\). In our BdG Hamiltonian, we find twice this 
number of zero eigenvalues ($2 n_{GM}=4$) providing a
consistency check.

On the antiferromagnetic side, with the \octahedral~order, all the symmetries are again broken.
At the same time, this ordering has \(\expval{Q^c}=0\) for any $c=x,y,z$ and hence \(n_{GM}=n_{BG}=3\)). In our BdG Hamiltonian, 
we find twice this number of zero eigenvalues ($2 n_{GM}=6$)
providing a consistency check.

Lastly, we find that at the pure chiral point, the number of zero modes scale with the system size if we start with \octahedral~or XYZ order, as is expected from the classical analysis as given in \cite{pitts2021order}. However, since we are
looking at fluctuations around potentially stable and unique ground states, we will consider the limit $J_3 \to 0^\pm$ and not the pure chiral point with $J_3=0$.

Since our analysis was done in real space, we get rid of the zero modes by adding small random diagonal terms to the Hamiltonian of the order \(\sim 10^{-10}\). Doing this also helps break degeneracies, which is essential for obtaining the BdG wavefunctions correctly \cite{Colpa1978}. After that is done, we end up in a basis \((\g_i,...,\g^{\dagger}_i,...)\) \textit{s.t.} \(\mh=2\sum_{i}\epsilon_i \left(\g^{\dagger}_i \g_i \right)+\epsilon_0\), where \(\epsilon_0\) is some constant energy shift which we can safely ignore. This new basis is related to the original basis through a similarity transformation, \(P\) (not a unitary transformation). Now we are interested in the expectation value, \(\a\equiv\frac{1}{N}\sum_{n,j}\expval{b^{\dagger}_{n,j} b_{n,j}}\). We can write \(\a\) as a linear combination of \(\expval{\g^{\dagger}_i \g_j}\,,\expval{ \g_i \g^{\dagger}_j}\,,\expval{ \g_i\g_j}, \expval{ \g^{\dagger}_i \g^{\dagger}_j}\) and some factors coming from \(P\). However, in the ground state, the only non-zero expectation is that of \(\expval{\g_i \g^{\dagger}_j}=\d_{ij}\), all the rest vanish.

\begin{figure}[ht]
    \centering
\includegraphics[scale=1]{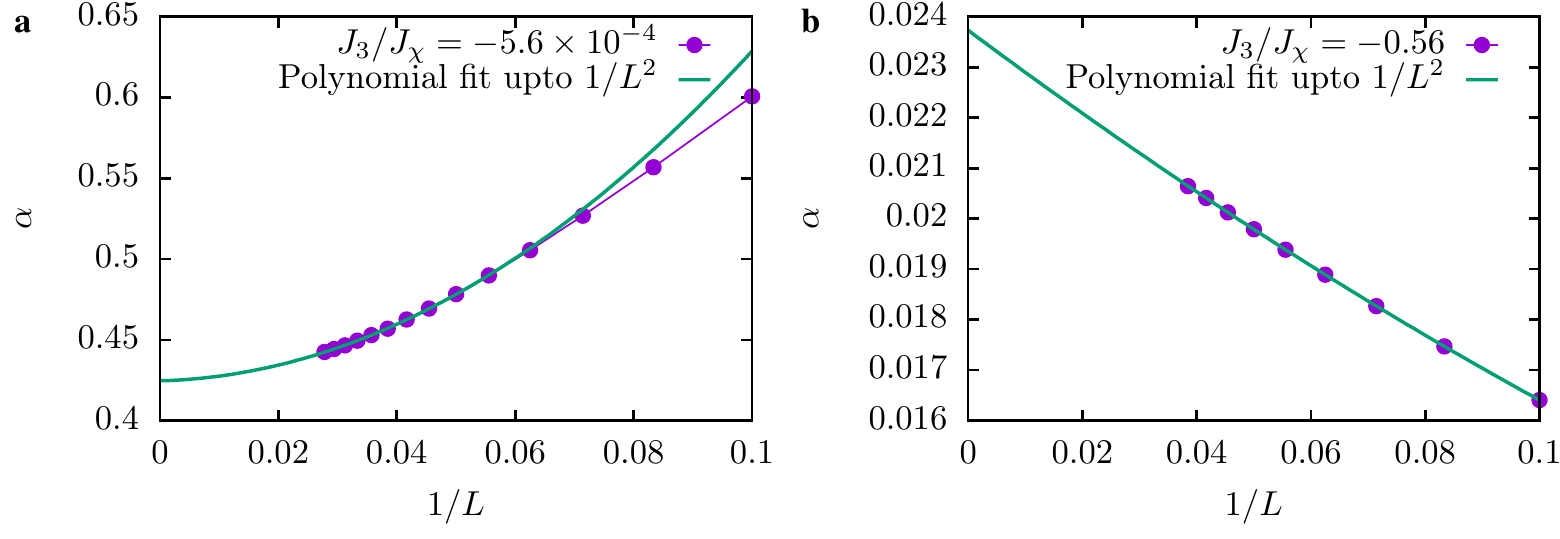}
    \caption{\(\a\) vs \(1/L\) for the ferromagnetic XYZ ordering, at different values of \(|J_3/J_{\c}|\).}
    \label{fig:kag_mag alpha vs L ferro}
\end{figure}

\begin{figure}[!ht]
    \centering
\includegraphics[scale=1]{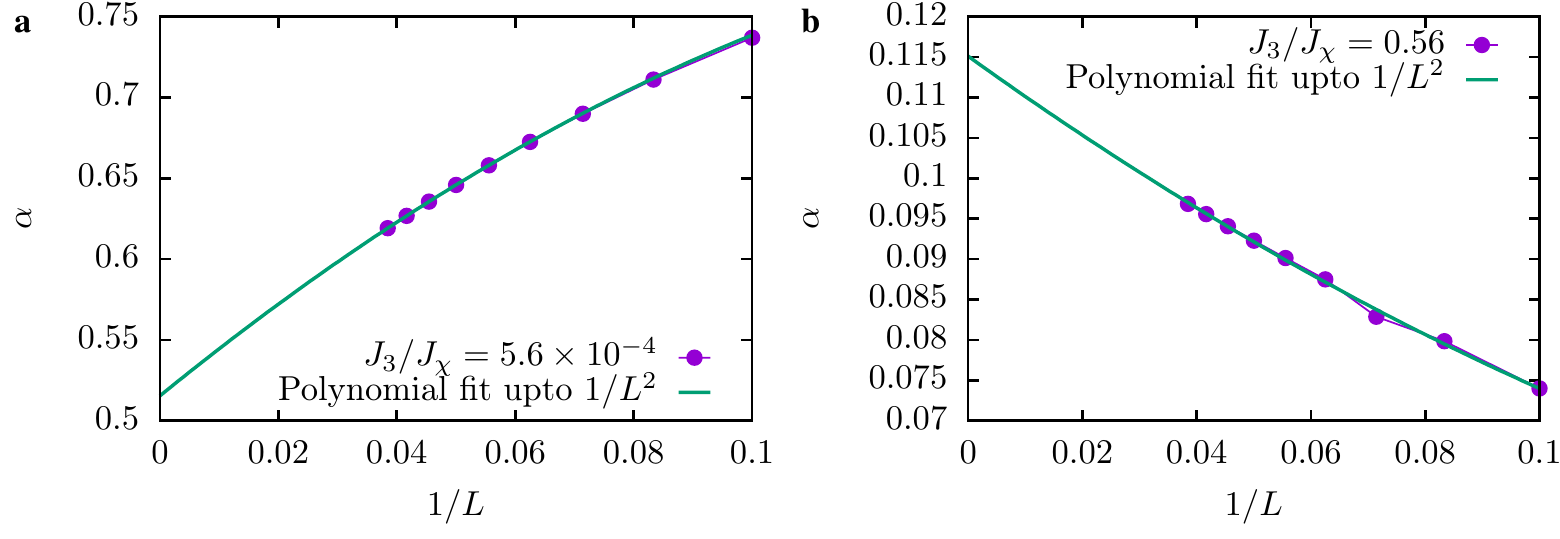}
    \caption{\(\a\) vs \(1/L\) for the anti-ferromagnetic \octahedral~ordering, at different values of \(|J_3/J_{\c}|\).}
    \label{fig:kag_mag alpha vs L antiferro}
\end{figure}

Doing this procedure for both the \octahedral~and the XYZ ordering gives us \(\a_{\pm}\) where \(\pm\) denotes the anti-ferromagnetic and the ferromagnetic cases. The finite size scaling of $\alpha_\pm(L)$ versus system size for $J_3 <0$ is shown in
Fig.~\ref{fig:kag_mag alpha vs L ferro} for two different values
of $J_3$. For small $|J_3|$, the fluctuations decrease with $L$
while for larger $|J_3|$ they increase with $L$, in both cases
extrapolating to a finite value as $L \to \infty$ when $J_3 \neq 0$.
Fig.~\ref{fig:kag_mag alpha vs L antiferro} shows similar finite
size scaling plots for $J_3 > 0$. From these plots, we extract the
thermodynamic limit value of $\alpha_\pm(J_3)$.


\begin{figure}[ht]
    \centering
\includegraphics[]{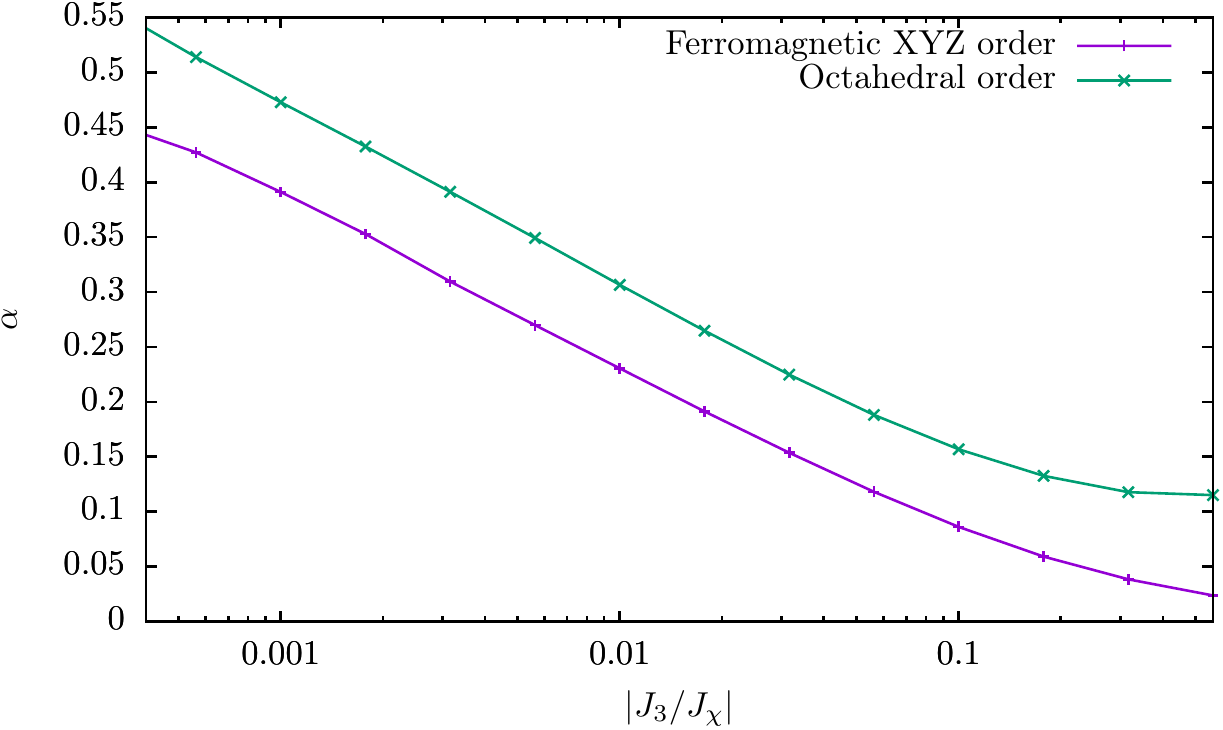}
    \caption{\(\a\equiv\frac{1}{N}\sum_{n,j}\expval{b^{\dagger}_{n,j} b_{n,j}}\) vs \(|J_3/J_{\c}|\) for the ferromagnetic XYZ ordering, and the Anti-ferromagnetic \octahedral~ordering.}
    \label{fig:kag_mag_correction}
\end{figure}

Fig.~\ref{fig:kag_mag_correction} shows the thermodynamic limit extrapolated value of $\alpha_\pm$ for both signs of $J_3$.
The $x$-axis is shown in a log scale to emphasize that \(\a_{\pm} = c_{\pm} \cdot \log\left(1/|J_3|\right)\) for small values of \(|J_3|\) (with \(J_{\c}\) fixed to 1). This logarithmic divergence as we approach the pure chiral Hamiltonian is consistent with the idea that long range order is completely melted away for all spins at \(J_3=0\). Furthermore, to estimate where the order melts for finite $S$, 
we have to come up with a definition \textit{s.t.} the value of \(|J_3|\) when \(\a_{\pm}\left(|J_3|\right) = f\cdot S\,,\:\:\: 0<f<1\), is the value where the ordering melts away for the spin, \(S\). The value of \(f\) which gives results matching semi-quantitatively with our ED data comes to about \(f\approx 0.4\).

\section{Parton mean-field theory}
\subsection{Operator expectation values}
We start by writing the spin operator in terms of the spin-\(1/2\) Schwinger fermion operators as
\begin{equation}
    \label{schwinger fermion}
    \boldsymbol{S}_i =\frac{1}{2} f_{i}^{\a\dagger}\boldsymbol{\s}_{\a\b}f_{i}^{\b}\,.
\end{equation}
In terms of these partons, the trial or the variational mean field Hamiltonian looks as follows
\begin{equation}
    \label{trial Hamiltonian}
    \begin{split}
    \mh_{trial} = - &\sum_{i,j} t_{ij}\sum_{\a}\left(e^{-\iota\phi_{ij}} f_{i}^{\a\dagger}f_{j}^{\a}+e^{\iota\phi_{ij}} f_{j}^{\a\dagger}f_{i}^{\a}\right) - \sum_{i} f_{i}^{\a\dagger}\left( \boldsymbol{\hat{b}}_i.\boldsymbol{\s}\right)_{\a\b} f_{i}^{\b}\,.
\end{split}
\end{equation}
Then, the expectation of the physical Hamiltonian is calculated in the ground state of such a variational Hamiltonian. Now since the variational Hamiltonian is just a free theory, the expectation of any operator can be written entirely in terms of two-point correlations through Wick's theorem. The two point correlations for the trial Hamiltonian are defined as
\begin{equation}
    \label{2-point corr}
    \chi_{ij}^{\a\b}\equiv \expval{f_{i}^{\a\dagger}f_{j}^{\b}}\,,\:\:\:\chi_{ii}^{\a\a}\equiv \expval{n_{i}^{\a}}\,.
\end{equation} 
Also note that, in general, \(\chi_{ij}^{\a\b}\) is not proportional to \(\d^{\a\b}\) because of the Weiss field which mixes the two spins.
Focusing on the Heisenberg like interaction term, we find that the relevant expectation value of the four-fermion operators looks like 
\begin{equation}
    \label{heis ferm expect}
    \begin{split}
    \expval{f_{i}^{\a\dagger}f_{i}^{\b}f_{j}^{\m\dagger}f_{j}^{\n}} & = \expval{f_{i}^{\a\dagger}f_{i}^{\b}}\expval{f_{j}^{\m\dagger}f_{j}^{\n}} - \expval{f_{i}^{\a\dagger}f_{j}^{\n}}\expval{f_{j}^{\m\dagger}f_{i}^{\b}} \\
    & = \chi_{ii}^{\a\b}.\chi_{jj}^{\m\n}-\chi_{ij}^{\a\n}.\chi_{ji}^{\m\b}\,,
    \end{split}
\end{equation}
where the relative \(-\) sign appears because of fermion anti-commutation relation. We therefore get that
\begin{equation}
    \label{heis expect}
    \begin{split}
        \expval{\boldsymbol{S}_i.\boldsymbol{S}_j} & =\frac{1}{4} \s^{a}_{\a\b}\s^{a}_{\m\n} \left[\chi_{ii}^{\a\b}.\chi_{jj}^{\m\n}-\chi_{ij}^{\a\n}.\chi_{ji}^{\m\b}\right]\\
        & = \frac{1}{4}\left(2\d_{\a\n}\d_{\b\m}-\d_{\a\b}\d_{\m\n}\right)\left[\chi_{ii}^{\a\b}.\chi_{jj}^{\m\n}-\chi_{ij}^{\a\n}.\chi_{ji}^{\m\b}\right]\\
          & =\frac{1}{4}\left(2\Tr(\chi_{ii}\:\chi_{jj})-\Tr(\chi_{ii}).\Tr(\chi_{jj})\right) - \left(2\Tr(\chi_{ij}).\Tr(\chi_{ji})-\Tr(\chi_{ij} \: \chi_{ji})\right)\,,
    \end{split}
\end{equation}
where we have used the completeness relation for Pauli matrices.

\subsection{Phase diagram and Chern numbers}

\begin{figure}[ht]
    \centering
\includegraphics[scale=1]{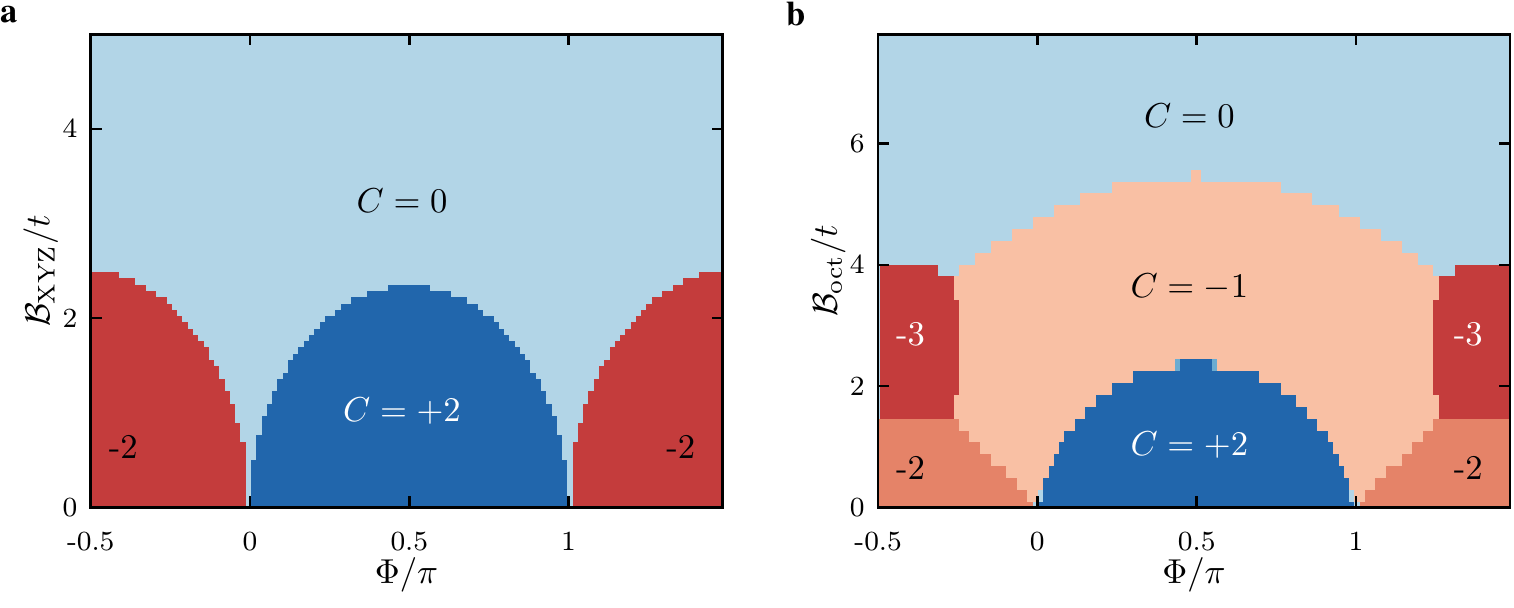}
    \caption{Chern number phase diagram for the ferromagnetic ({\bf a}) and the anti ferromagnetic ({\bf b}) cases obtained using parton mean-field theory.}
    \label{fig:kag cherns}
\end{figure}

The Chern number \cite{Fukui:2005wr} phase diagrams are obtained using Eq.~\eqref{trial Hamiltonian} with the hopping being restricted to nearest neighbours at unit strength. The flux pattern is varied \textit{s.t.} there is a flux of \(\Phi\) through each up and down-triangle, and a flux of \(\pi-2\Phi\) through each hexagon of the \kagome~lattice. Lastly, an ordering is chosen for the Weiss fields, but its strength, \(\mathcal{B}\), is varied. The ordering is chosen from our mean field results, when we optimize the total energy for the Chern insulator after switching on bow-tie Heisenberg interactions. In the main manuscript, we have shown the Chern number evolution in Fig.~2 at fixed $\Phi/\pi=0.5$, fixed by the effective Hofstadter model, 
and the spontaneously induced ${\cal B}$.

For the antiferromagnetic case, the ordering turns out to be \octahedral~after a critical value of \(J_3=J_3^{\mathrm{oct}}>0\), while for the ferromagnetic case,  below a critical negative value of \(J_3=J_3^{\mathrm{XYZ}}<0\), the ordering is that of a squished XYZ state, which interpolates between perfect XYZ order and the \(Q=0\),  \(120^\circ\) coplanar state. 
We are calling this the ``XYZ'' state since it is an umbrella order
with the same symmetries as the XYZ state. 

\section{Exact Diagonalization}
The unit cells we have used for performing exact diagonalizations (ED) are shown in Fig. 4{\bf f} in the main paper. On a regular lattice every site participates in two unique bow-tie couplings. In order to compare results for the different size unit cells, care has to be taken when implementing the periodic boundary conditions on the $N=12$ and $N=24$ rectangular clusters. For the $N=12$ cluster, all bow-tie bonds can connect both ways around the torus and are therefore counted twice. For the $N=24$ cluster, there are 8 such bow-tie bonds that connect both ways around the small circumference of the torus and are therefore counted twice, while the remaining bow-tie bonds are only counted once. For the rhombic $N=24Rh$ unit cell as well as for the $N=36$ cell, all the $2N$ bow-tie bonds are uniquely defined with periodic boundary conditions and are only counted once.
\subsection{Phase Diagram}

\begin{figure}[ht]
     \centering
\includegraphics[width=0.495\textwidth]{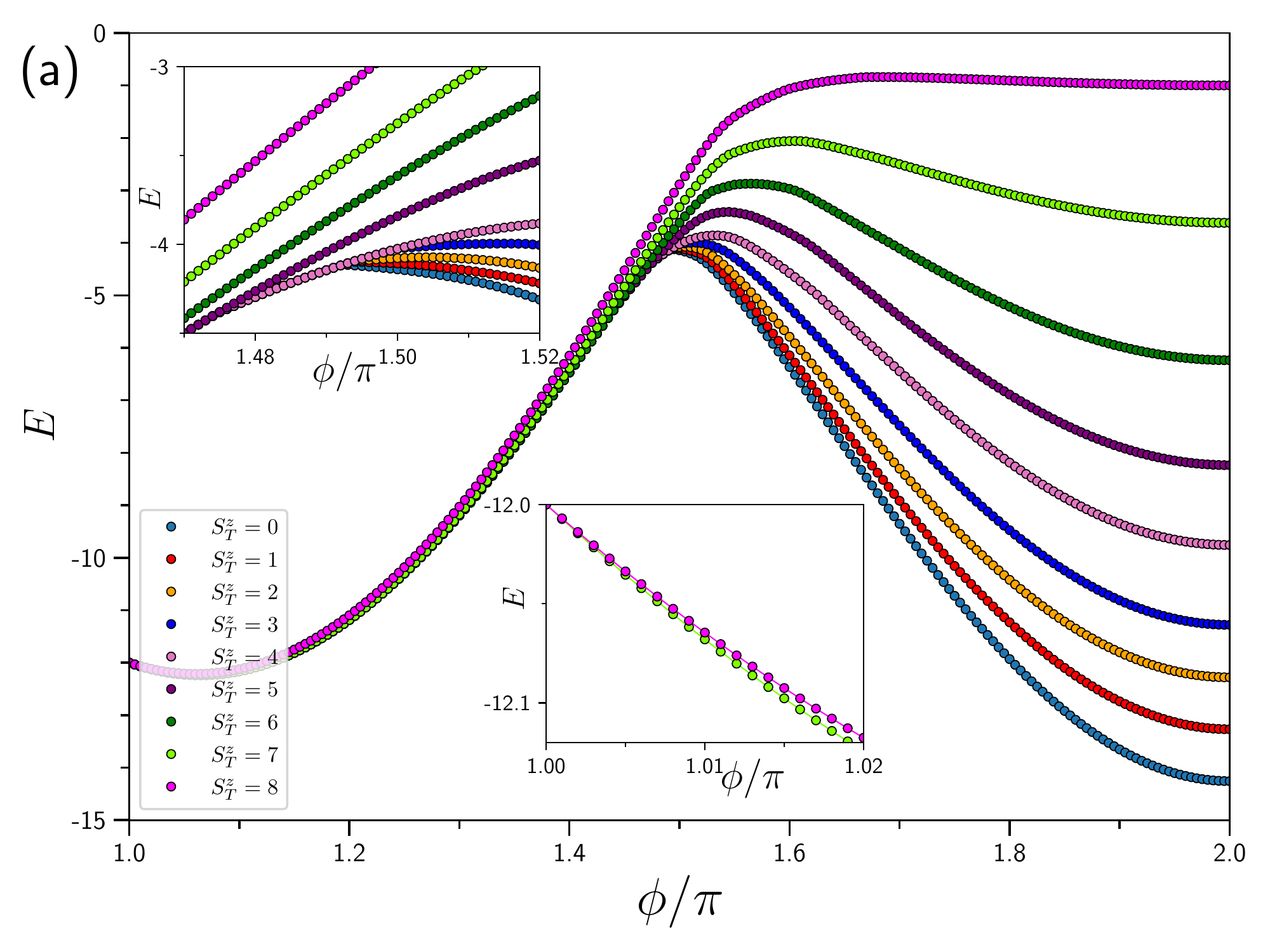}
\includegraphics[width=0.495\textwidth]{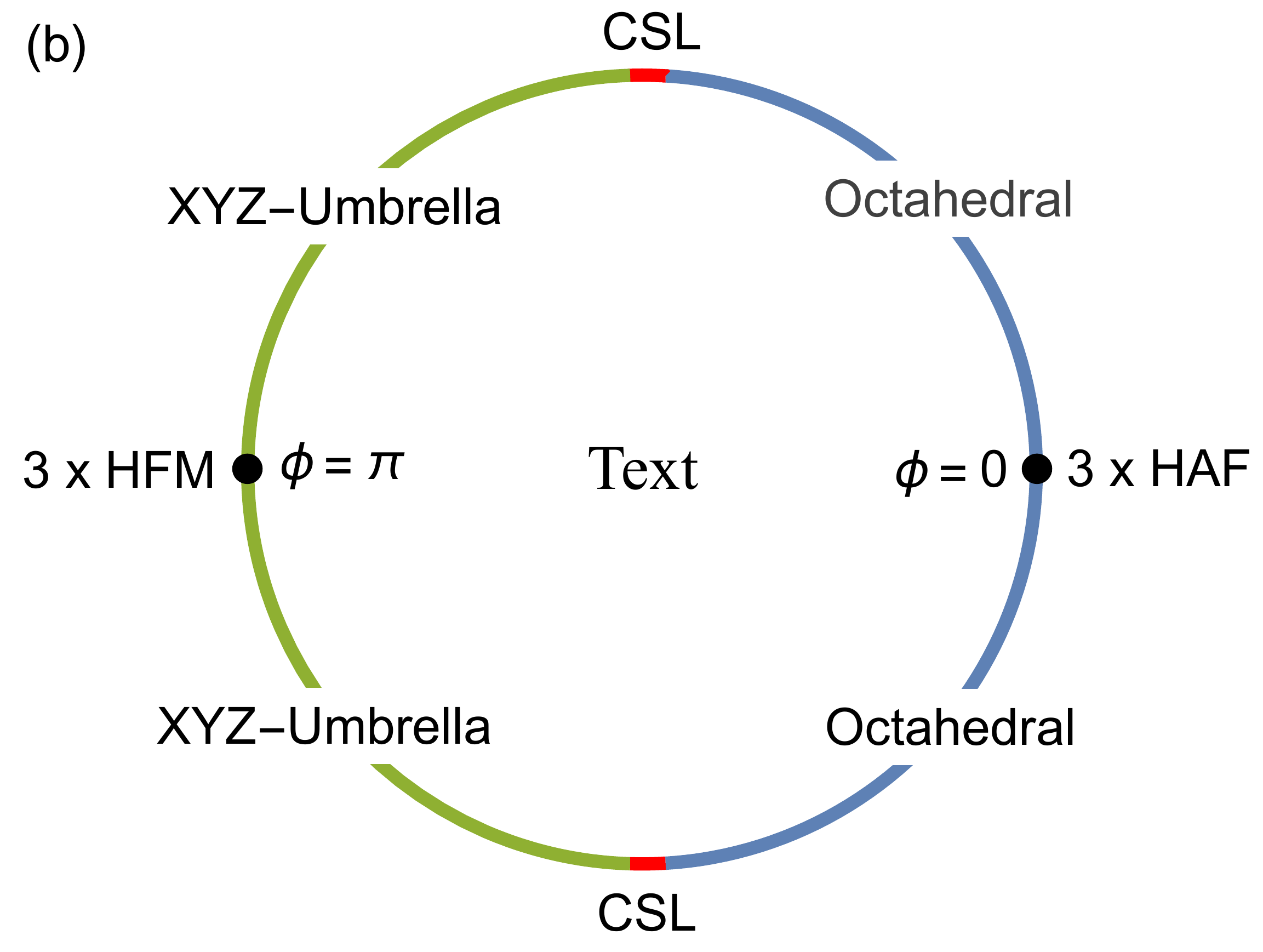}
        \caption{(a) Total energy versus $\phi/\pi$ for the $N=24Rh$ unit cell. The insets show close up behavior in the vicinity of $\phi=\pi$ and $\phi=3\pi/2$. Note that the $S^z_T=8$ state is not degenerate with $S_T^z$ until $\phi=\pi$. (b) Phase diagram as a function of $\phi$. The extent of the CSL phase is estimated from the $N=36$ results in the main paper. For $\phi=0,\pi$ the system decouples into 3 separate lattices.}
        \label{fig:PhaseDiagram}
\end{figure}

To gain a more complete understanding of the phase diagram we parameterize the couplings in $H_{\rm spin}$ in the following way:
\begin{equation}
      H_{\phi} = \sin(\phi)\sum_{\bigtriangleup,\bigtriangledown}
    \vS_i\cdot \vS_j \times \vS_k 
    + \cos(\phi)\sum_{\btjnbrR~~~\btjnbrL} \vS_i\cdot \vS_j.
\end{equation}
With this parameterization we can explore the full phase diagram of the model, reaching the limits of $J_\chi=0\ (\phi=0,\pi)$ described by the 3 lattice toy model from section~\ref{sec:toy} at $\phi=\pi$.
We first note that this model is invariant under $\phi\to 2\pi-\phi$ which leaves 
$\cos(\phi)$ and thereby the nearest neighbor Heisenberg term unchanged but changes the sign of the chiral term. However, it is easy to see that the chiral interaction is independent of the sign implying the invariance. Strictly speaking we therefore only need to consider $\phi\in (0,\pi)$ or equivalently $\phi\in (\pi,2\pi)$.

The lowest energy for each of the sectors $S_T^z=0\ldots 8$ for the $N=24Rh$ lattice are shown in Fig.~\ref{fig:PhaseDiagram}(a) for $\phi\in (\pi,2\pi)$
with the insets showing detailed behavior close to $\phi=\pi $ and $3\pi/2$. The spectrum is mostly dominated by the nearest neighbor Heisenberg coupling leaving only a small region close to $\phi=\pi/2,3\pi/2$ (estimated from the 36-site results in the main paper) for the CSL (see upper inset in Fig.~\ref{fig:PhaseDiagram}(a)).
On the ferromagnetic side, $\pi/2<\phi<3\pi/2$, and close to $\phi=\pi$, the second inset shows that a gap remains to the $S^z_T=8$ state until the completely decoupled lattices are reached at $\phi=\pi$. This is consistent with the prediction of $S_T=7$ for the ground-state in this limit obtained from the toy model (section~\ref{sec:toy}). In a similar manner we expect
that we reach 3 decoupled anti-ferromagnetic nearest neighbor lattices only precisely at $\phi=0$ with the preceding phase being characterized by \octahedral~ordering. A sketch of the expected phase-diagram is shown in Fig.~\ref{fig:PhaseDiagram}(b).

\subsection{Spectrum with quantum numbers}
We restrict a more complete symmetry analysis, including more quantum numbers than the $S^z_T$ used in the main paper, to the small $N=12$ cluster ($2\times2$ \kagome~unit cells).

\begin{figure}
\centering
\includegraphics[scale=1]{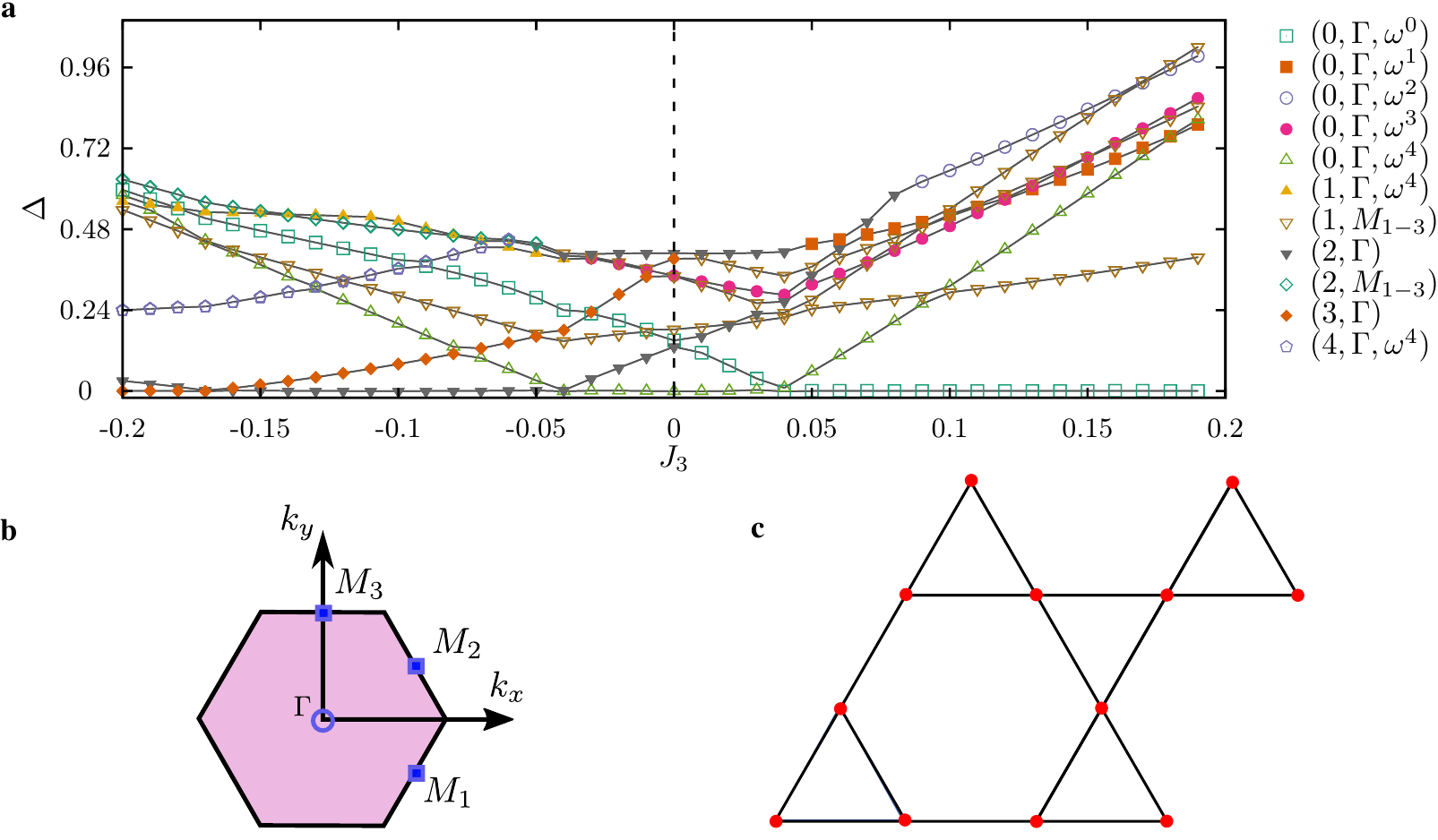}
    \caption{{\bf a} Energies of eigenstates, relative to the ground state energy, arranged according to the quantum numbers -- total spin, momenta ($\Gamma$, $M_1$, $M_2$, $M_3$), and \(C_6\) rotation (about hexagon centre) eigenvalues at the $\Gamma$ point
    (with \(\w=e^{\i\pi/3}\)). {\bf b} The hexagonal Brillouin zone showing the special momentum points related to lattice translational symmetry. {\bf c} The $12$-site cluster used in the ED calculations, which is also shown in Fig.4 in the
    main manuscript.}  
    \label{fig:kagome ED}
\end{figure}

For $J_\chi\!=\! 1$ and $J_3\!=\!0$,
i.e., the pure chiral model, the
ground state energy per site is $\approx\!-0.186 J_\chi$. This
ground state is a spin singlet at the $\Gamma$-point 
$\bk=(0,0)$, but it has a non-trivial $C_6$ eigenvalue 
for $2\pi/6$
rotations about the \kagome~hexagon centre, with
$\lambda^{(1)}_0(C_6)=e^{i 4\pi/3}$. The next singlet in the
spectrum is also a $\Gamma$-singlet but with rotation eigenvalue $\lambda^{(2)}_0(C_6)=1$. We expect these two
singlets, which are separated by a gap $\sim\!0.151 J_\chi$
on our small system size, 
to become the two topologically degenerate levels
of the CSL in a large system; indeed, these rotation eigenvalues
are consistent with what we would obtain from the $S$ and $T$ 
matrices for the Abelian anyons (semions) of the CSL \cite{Vidal2013}. 
For the $N=12$ cluster the second singlet is higher in energy than the first triplet at $J_3=0$. For the $N=36$ cluster in Fig 4a in the main paper it appears below the triplet at $J_3=0$ consistent with our expectation that it becomes degenerate with the ground-state singlet in the thermodynamic limit.

With increasing bow-tie exchange $J_3$, we find that
the energy of one of these singlets decreases while the
other singlet drifts up in energy. At the same time, a set of
triplets, with momenta at the $M$-points of the hexagonal
Brillouin zone (BZ), come down in energy. We tentatively 
identify the point where the
upward drifting 
singlet crosses the downward moving $M$-triplet, which occurs
at $J_3/J_\chi \approx 0.1$, as
a CSL to magnetic order transition point. This is in qualitative agreement with the estimate of $J_3/J_\chi \approx 0.06$ for the $N=36$ cluster discussed in the main paper.

\begin{figure}
     \centering
\includegraphics[width=0.495\textwidth]{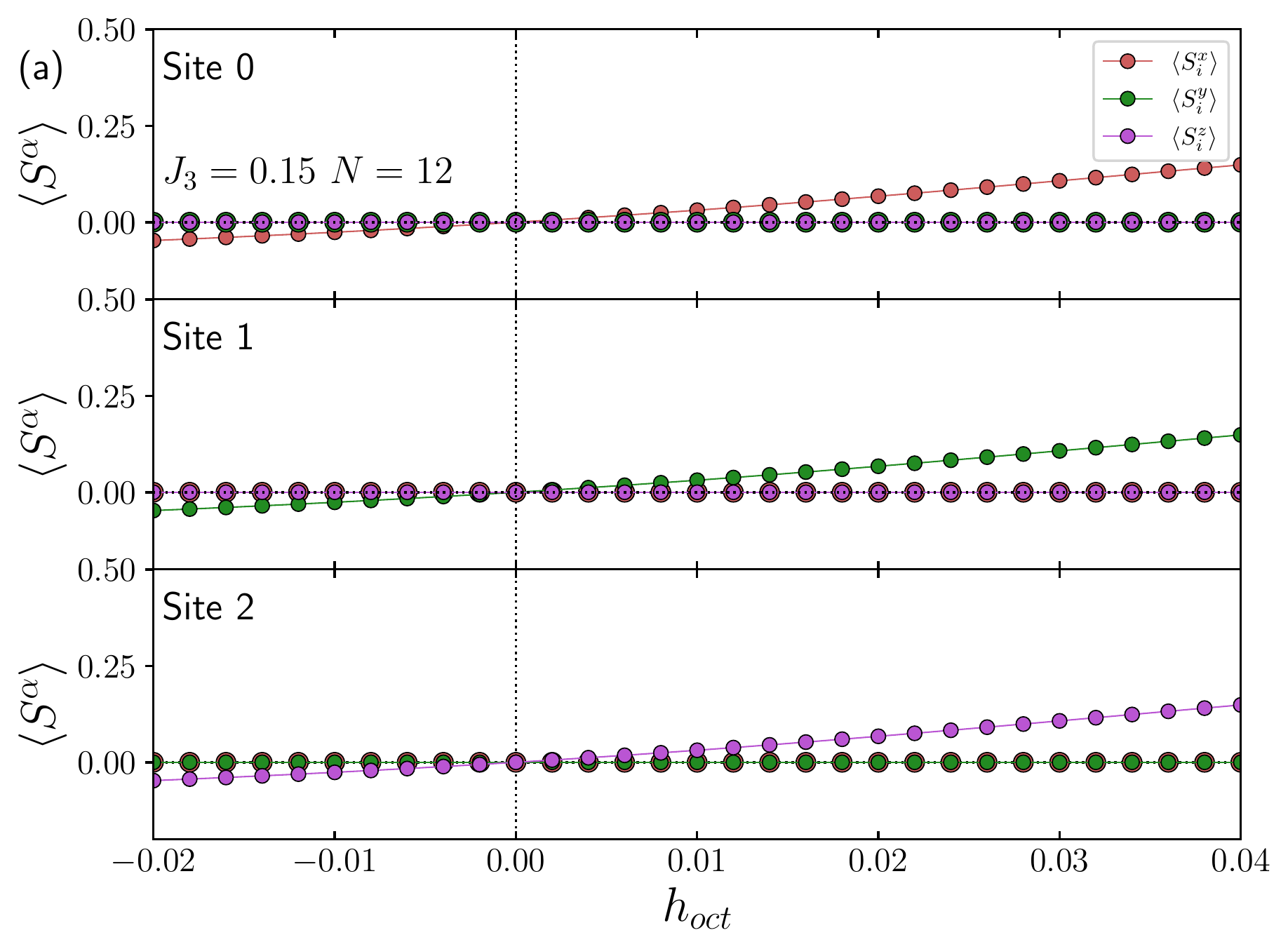}
\includegraphics[width=0.495\textwidth]{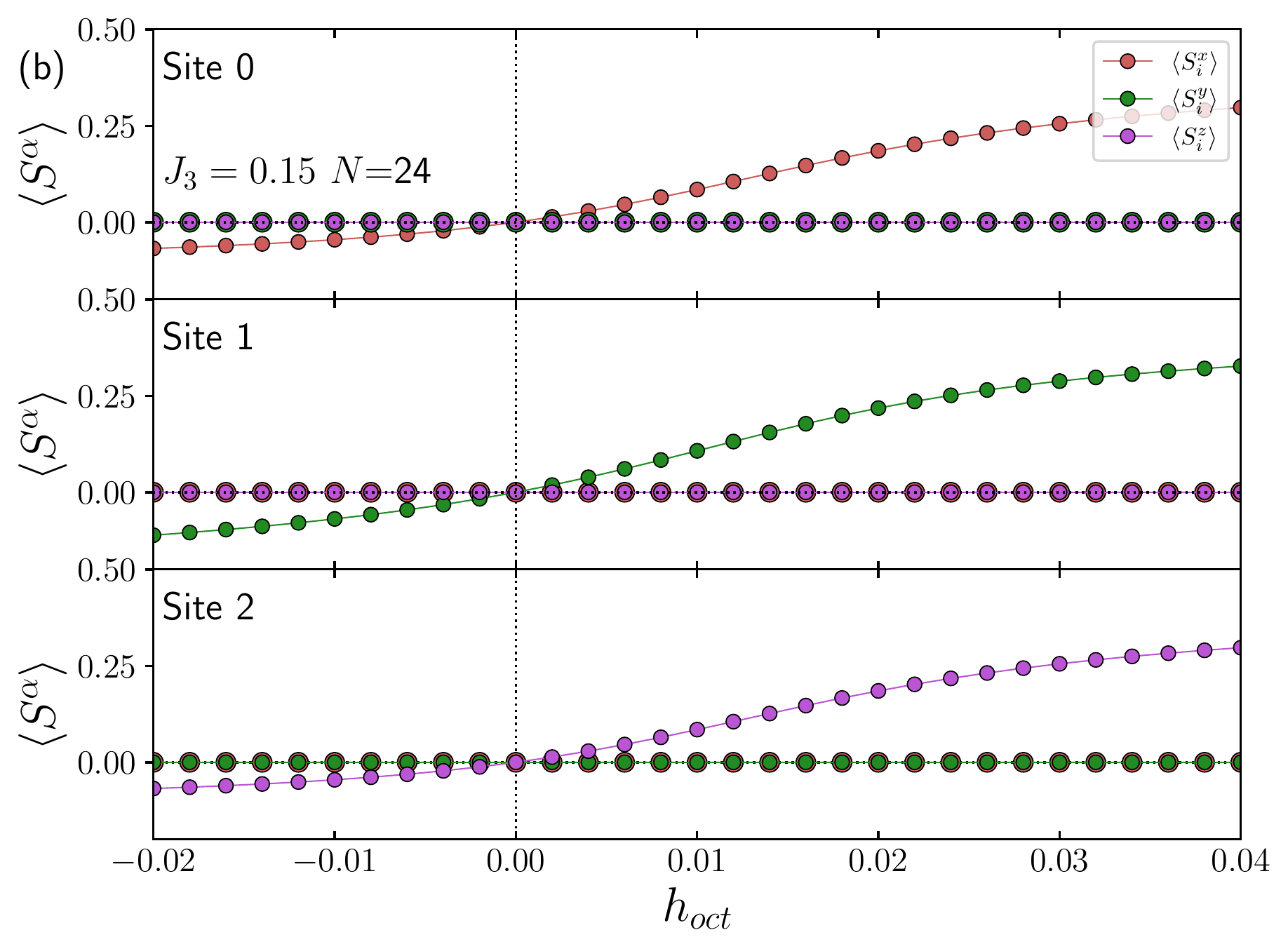}
\includegraphics[width=0.495\textwidth]{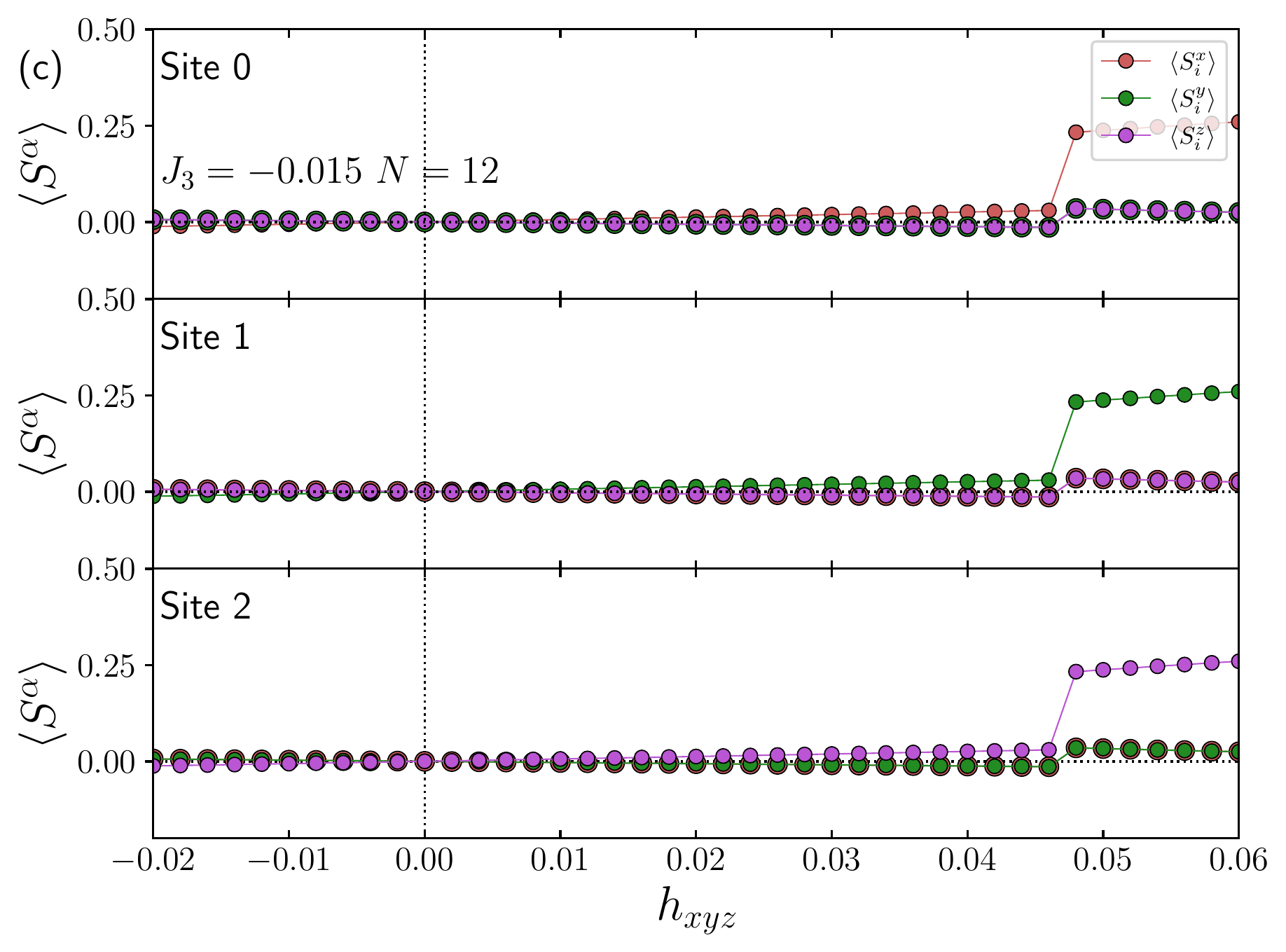}
\includegraphics[width=0.495\textwidth]{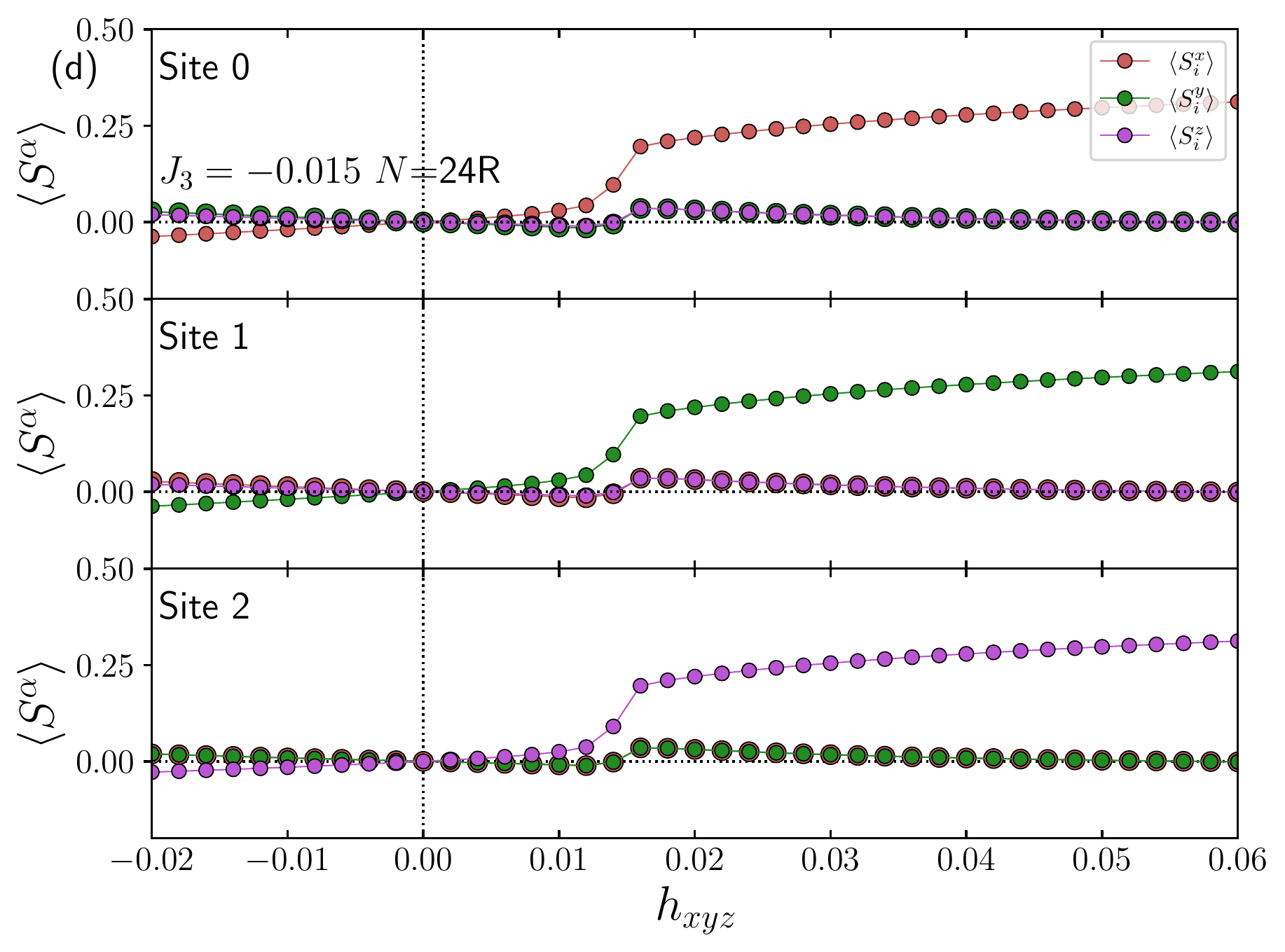}
        \caption{$\langle S^a\rangle$ on subsequent sites around a triangle for the $N=12$ unit cell at $J_3=0.15$ for (a) $N=12$ and (b) $N=24$ versus applied field \hoct~and, for $J_3=-0.015$ for (c) $N=12$ and (d) $N=24Rh$ versus applied field \hxyz.}
        \label{fig:suscept}
\end{figure}
\subsection{Response to global \hxyz~and \hoct~fields}
As discussed in the main paper, we consider the response of the system to global Zeeman fields \hxyz~and \hoct~inducing the XYZ-umbrella and \octahedral~orderings. In Fig.~\ref{fig:suscept} we show additional
results. Fig.~\ref{fig:suscept}(a),(b) illustrate the behavior of the system at $J3=0.15$ where \octahedral~ordering is present when the system size is increased from (a) $N=12$ to (b) $N=24$. (Fig.~\ref{fig:suscept}(a) is identical to Fig. 4c in the main paper.) Since the field is applied throughout the lattice we an limit our analysis to a single triangle with adjacent sites labelled 0,1,2 (anti clock-wise). As is clearly observed in Fig.~\ref{fig:suscept}(a),(b) the response is {\it significantly} stronger for $N=24$ and if a susceptibility, $\partial \langle S^\alpha \rangle/\partial \mhoct$ with respect to \hoct~is defined we would expect it to {\it diverge} with $N$ at zero applied field ($\mhoct=0$), consistent with the presence of \octahedral~ordering. 

It is also instructive to analyze the response to XYZ-umbrella ordering within the CSL phase. From the results presented in the main paper we know that for $J_3 \lesssim -0.03J_\chi$ the ordering spontaneously jumps to large values for any finite field. In Fig.~\ref{fig:suscept}(c),(d) we show results at $J_3=-0.015J\chi$
within the CSL phase for (c) $N=12$ and (d) $N=24Rh$ versus applied field \hxyz. The onset is again abrupt but now appears at {\it finite} field strengths. Since we do not expect the spin gap to close completely in the CSL phase we expect that a finite field will always be needed to induce the XYZ-umbrella ordering even though the gap decreases noticeable between $N=12$ and $N=24Rh$ as reflected in the shift in the onset of ordering from $ \mhxyz\approx 0.048$ to $\mhxyz\approx 0.014$. 

\begin{figure}[ht]
    \centering
\includegraphics[scale=1]{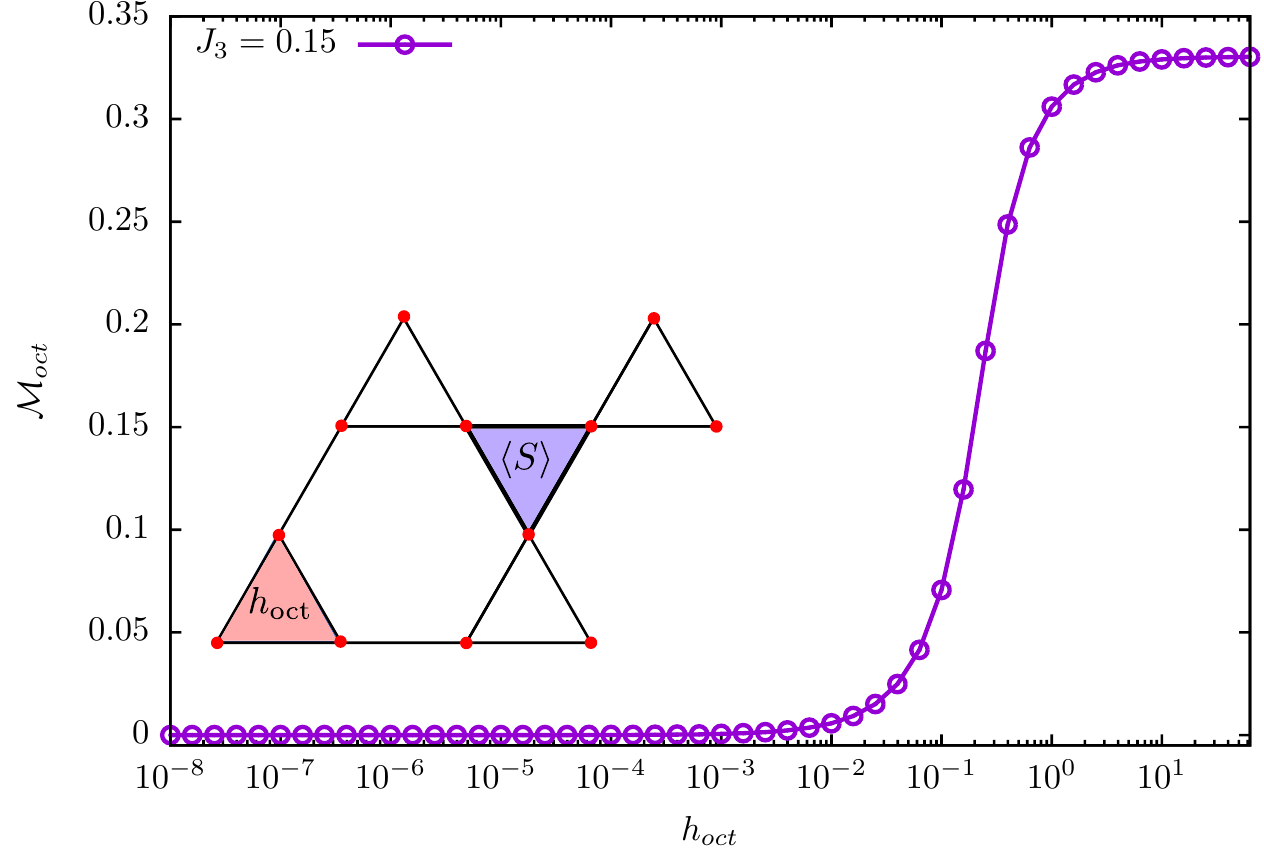}
    \caption{The overlap of induced magnetization at \(J_{\c}=1\,,J_3=0.15\) with the expected \octahedral~pattern on the blue triangle, different from the one where a Zeeman field is applied (orange triangle), with strength \(\mhoct\) for $N=12$ versus \hoct.}
    \label{fig:kag induced magoct}
\end{figure}

\subsection{Induced magnetization from  local \hoct~and \hxyz~fields}
As explained in the main paper, it is instructive to examine
the nature of magnetic ordering induced by a local Zeeman field applied around {\it a single}
triangle by introducing a term in the spin
Hamiltonian of the form $-\mhoct \left(S^z_0 + S^x_1 + S_2^y\right)$. The field  then
points along $\hat{z}$, $\hat{x}$, and \(\hat{y}\) respectively at 
the three adjacent sites $i=0,1,2$ around the triangle and the response can be studied throughout the lattice as the field is varied. The introduction of \hoct~completely break all
the spin and lattice symmetries, and will mix the 
ground state with the low-lying states. Note that there is no need to introduce a field on {\it more} than 3 sites to uniquely induce the \octahedral~ordering. The introduction of a local \hxyz~field is identical in form.

We have computed the resulting induced moments at all sites $\la \vec S_i \ra$ on the cluster. We then choose a triangle furthest away from the one where the Zeeman field is applied, and calculate the overlap, \moct~of the induced magnetization on this triangle with the expected \octahedral~pattern, and plot it as a function of the Zeeman field strength. This is shown in Fig.~\ref{fig:kag induced magoct}.

\begin{figure}[ht]
    \centering
\includegraphics[width=0.495\textwidth]{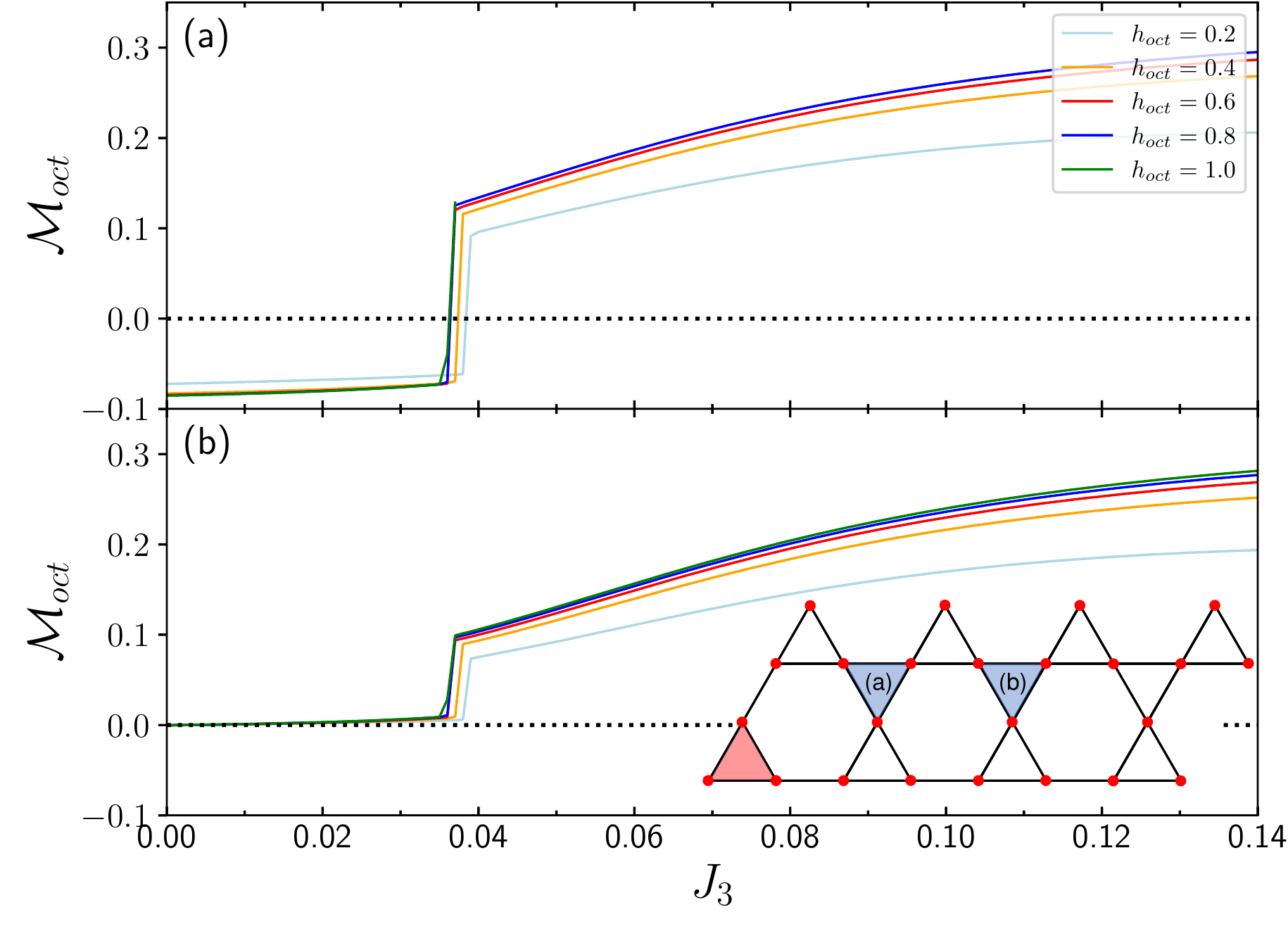}
\includegraphics[width=0.495\textwidth]{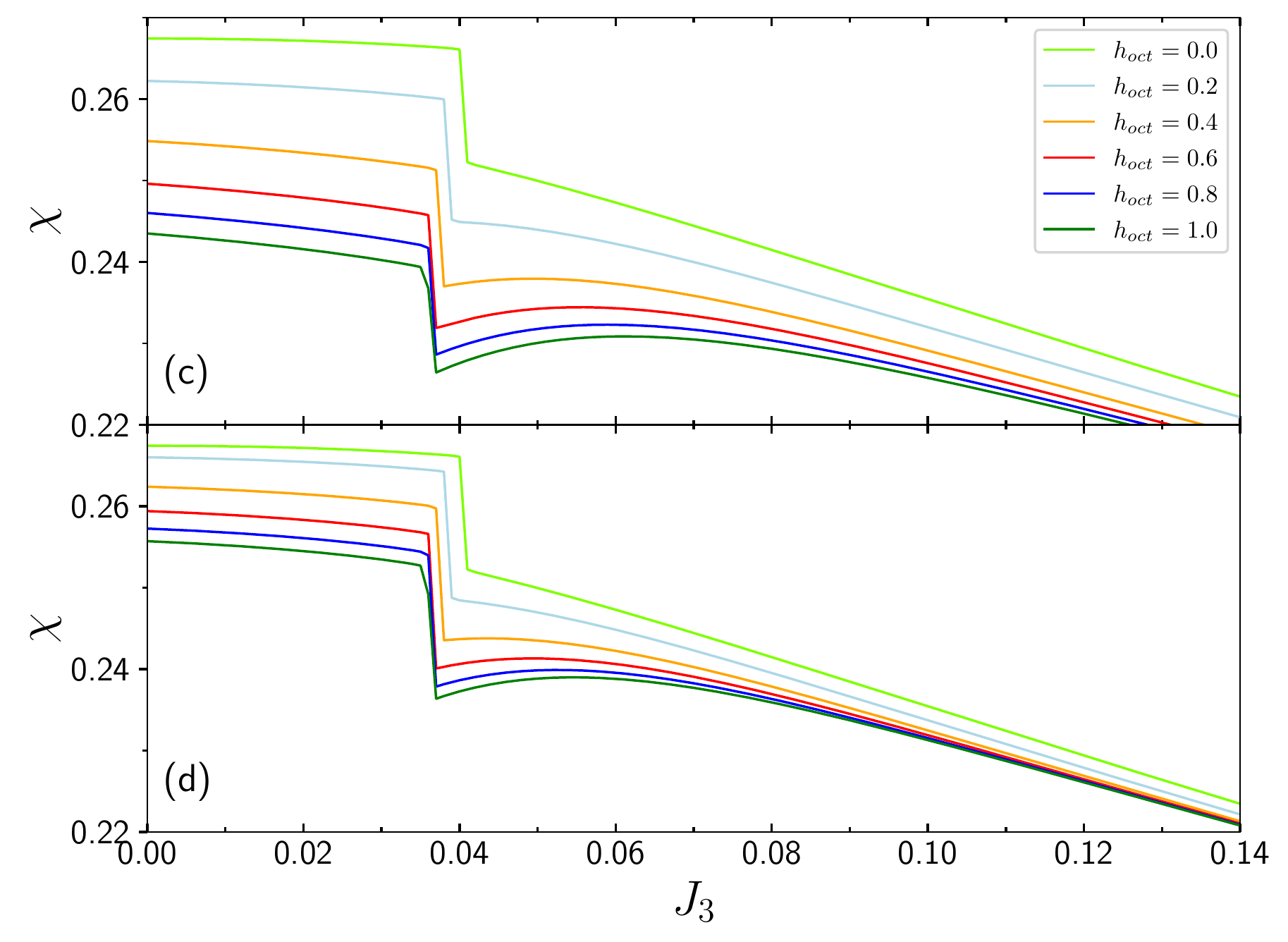}
    \caption{(a), (b) The overlap, \moct~of induced magnetization with the expected \octahedral~pattern versus $J_3$ for a range of field strengths $\mhoct=0.2,0.4,0.6,0.8,1.0$ shown for two different triangles (blue shaded). The field is applied on the red triangle. A regular $N=24$ unit lattice was used.
    (c), (d) The scalar chirality $\chi=\langle \vec S_i\cdot(\vec S_j\times \vec S_k)\rangle$ around the same two triangles as shown in panels (a), (b).}
    \label{fig:moct}
\end{figure}
To further explore how the \octahedral~ordering is induced we have repeated the calculation of \moct~of the $N=24$ site lattice as a function of $J_3$ (with $J_\chi=-1$) for a range of field strengths $\mhoct=0.2,0.4,0.6,0.8,1.0$. Our results are shown in Fig.~\ref{fig:moct}. The field is again applied only at a single triangle shown in red in Fig.~\ref{fig:moct} while \moct~is calculated on two different triangles shown in blue. Panel (b) corresponds to the triangle furthest away from the red triangle. Clearly the \octahedral~pattern appears rapidly at even modest applied fields for sufficiently large $J_3$. For $J_3\lesssim 0.04$ the CSL phase is clearly visible and the \octahedral~order is absent. For this $N=24$ the transition between the CSL and the \octahedral~ordered phase appears first order at finite field.

For comparison, it is instructive to study the behavior of scalar chirality $\chi=\langle \vec S_i\cdot(\vec S_j\times \vec S_k)\rangle$ on the same triangles as $J_3$ is varied. This is shown in Fig.~\ref{fig:moct}(c),(d). The transition between the CSL and \octahedral~phase is again clearly visible. For $\mhoct=0$ $\chi$ is uniform among all triangles in the lattice. Note that, as the applied \octahedral~field \hoct~is increased the value of the scalar chirality, $\chi$, decreases towards its maximal classical value of $1/8$.

\subsection{Transition from NN \kagome~Heisenberg AF to Chiral Spin Liquid}

\begin{figure}
\centering
\includegraphics[scale=0.6]{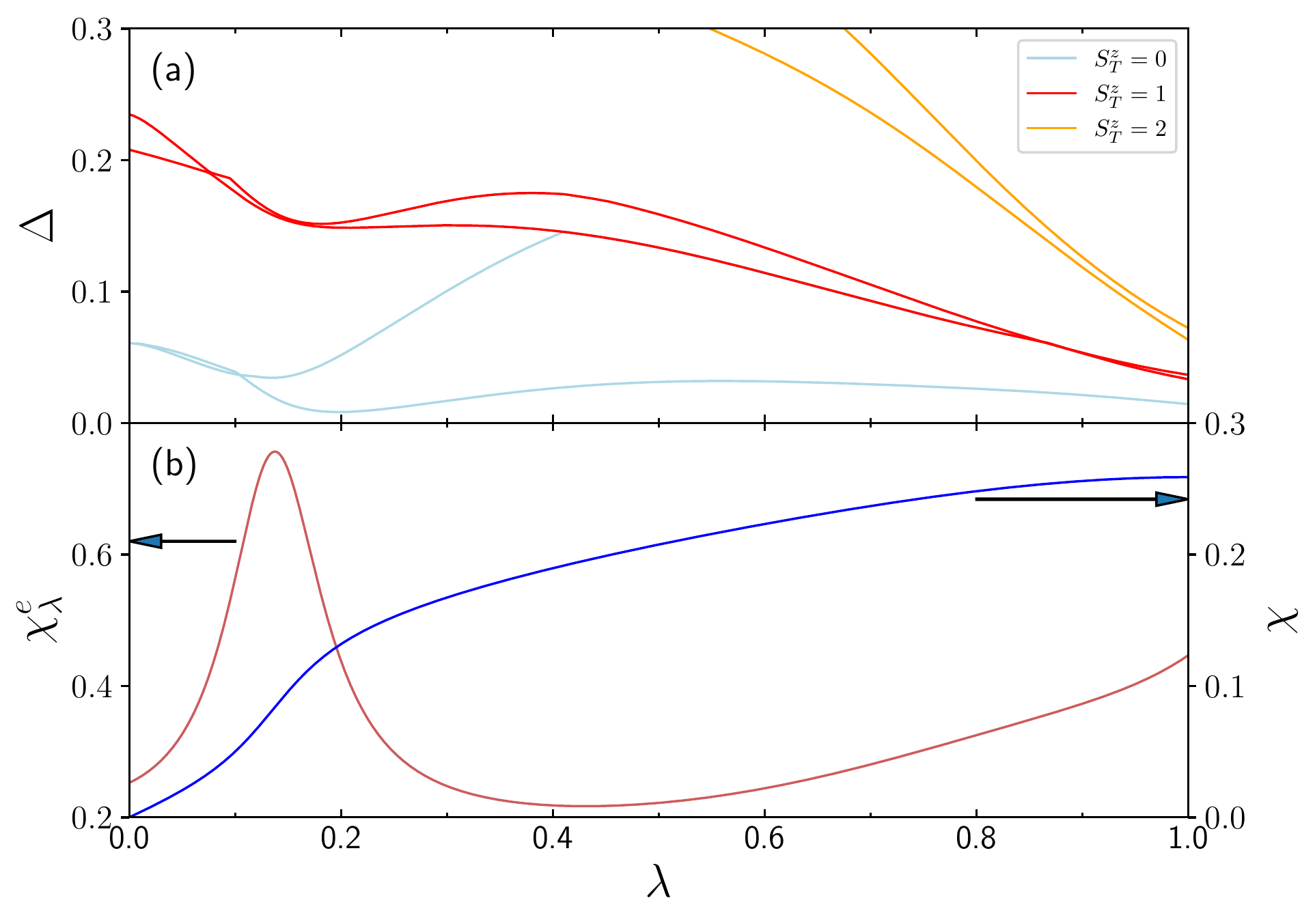}
    \caption{(a) The gap, $\Delta$, to the lowest lying $S^z_T=0,1,2$ states versus $\lambda$ for the $N=24Rh$ cluster. (b) The energy susceptibility, $\chi^e_\lambda$, and scalar chirality $\chi=\langle \vec S_i\cdot(\vec S_j\times\vec S_k)\rangle$, versus $\lambda$ indicating a second order transition in the vicinity of $\lambda_c\approx 0.14$}  
    \label{fig:kag2csl}
\end{figure}

It is expected~\cite{Bauer2014} that the chiral spin liquid at $J_\chi=1,\ J_3=0$ is distinguishable from the phase of the nearest neighbor Heisenberg antiferromagnet on the \kagome~lattice. To illustrate this we consider the following combined model extrapolating between the two limits:
\begin{equation}
\label{eq:lambdaham}
    H_\lambda=-\lambda J_\chi\sum_{\bigtriangleup,\bigtriangledown}
    \vS_i\cdot \vS_j \times \vS_k 
    + (1-\lambda)J\sum_{<i,j>} \vS_i\cdot \vS_j.
\end{equation}
Here, $J$ is the usual nearest neighbor (NN) coupling between sites on the \kagome~lattice. We use the $N=24Rh$ unit cell to study the phase-diagram of $H_\lambda$ as $\lambda$ is varied between $\lambda=0$ (the pure NN HAF \kagome~model) and 
$\lambda=1$ the purely chiral model. A convenient way of detecting quantum phase transitions is by analysing the ground-state energy susceptibility:
\begin{equation}
    \chi^e_\lambda = -\frac{\partial^2 e_0(\lambda)}{\partial\lambda^2},
\end{equation}
where $e_0$ is the ground-state energy per spin. It can be shown~\cite{Alet2010} that at a quantum critical point (QPT) $\chi^e_\lambda\sim L^{2/\nu-(d+z)}$, where $L$ is the linear size of the system. Hence, as long as $2/(d+z)>\nu$ we expect to see a divergence in $\chi_\lambda^e$ with $N,L$ at the QPT. Our results for the gap to the lowest lying states for $S^z_T=0,1,2$ are in Fig.~\ref{fig:kag2csl}(a) and for $\chi^e_\lambda$ in Fig.~\ref{fig:kag2csl}(b) along with the scalar chirality $\chi$. At $\lambda=0$ the spectrum is dominated by low-lying singlets with the gaps to $S_T^z=1,2$ states rapidly decreasing with $\lambda$. Close to $\lambda_c\sim 0.14$ a significant peak in $\chi^e_\lambda$ is visible consistent with a second order phase transition. At the same time $\chi$ increases from zero at $\lambda=0$ to $0.259$ for the CSL at $\lambda=1$ with the most rapid increase in the region around $\lambda_c\sim 0.14$. The saturation value of $\chi=0.259$ is slightly lower for the $N=24Rh$ cluster as compared to the value of $\chi=0.267$ ($J_3=0, \mhoct=0)$ for the $N=24$ cluster shown in Fig.~\ref{fig:moct}(c), (d).

\subsection{Toy Model for Ferromagnetic bow-tie interaction}\label{sec:toy}

\begin{figure}[ht]
    \centering
\includegraphics[scale=1]{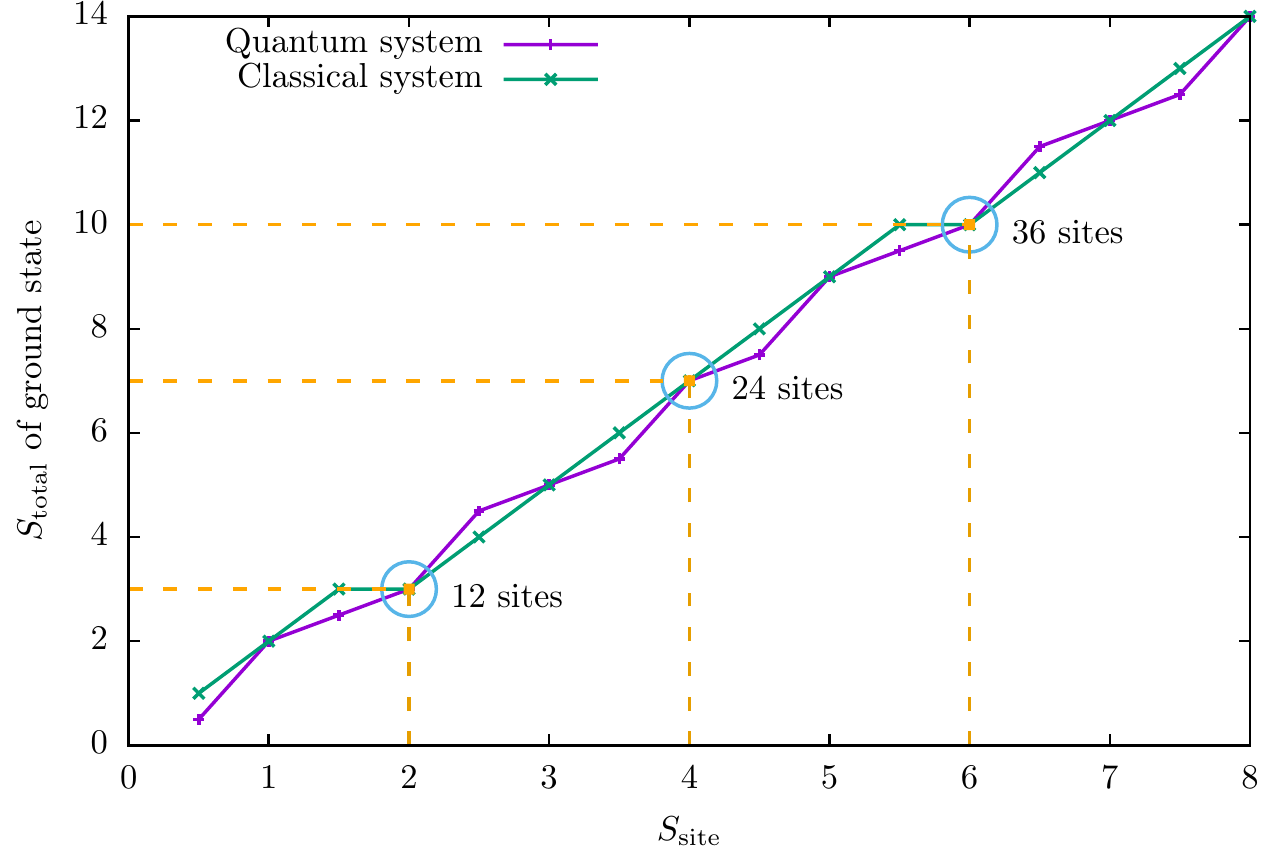}
    \caption{Total spin of the ground state vs the spin on each site of the triangle for both a quantum and a classical system with \(\mh=\mh_{\c}+ \mh_{bt}\).}
    \label{fig:tri toy model}
\end{figure}

The ferromagnetic bow-tie Heisenberg interaction on the \kagome~lattice for spin $S=1/2$, 
when added to a Chiral interaction, can be explained through a simple toy model. The idea is that at \(J_3/J_{\c}\rightarrow -\infty\), the \kagome~lattice is broken up into three ferromagnetic square lattices. On these square lattices, each spin point to the same direction. Hence each square sub-lattice hosts a total spin state, \(S_{\mathrm{square}}=N/6\), where \(N\) is the total number of sites on the \kagome~(\(N/3\) being the number of sites on each square sub-lattice, multiplied by a spin \(1/2\) on each site). 

Now, when the chiral term is switched on, effectively, each triangle on the \kagome~will act as if these large spins are interacting through a chiral like term. So we can find the ground state of the full problem by just solving a single triangle with a large spin, \(S=S_{\mathrm{square}}\), 
present on each site. Doing so, we find the total spin of the ground state, which matches with the ED results on the $N=12,24,36$ site systems.

Furthermore, it can be seen that the resulting total spin of the ground state actually matches with the predicted classical value of spin. This is so as the classical ground state of the chiral plus a ferromagnetic bow-tie Heisenberg is just the XYZ order. For the XYZ order, the total classical spin length on each triangle is just \(\sqrt{3}\cdot S\), and the quantum results match with this (upto taking a nearest integer value), as shown in fig.(\ref{fig:tri toy model}).

\ifdefined\makeSM{}
\else
\bibliography{ref}
\end{document}
 \fi
\end{document}